\documentclass[
aps,
rmp,
reprint,
amsmath,
amssymb,
graphicx,
longbibliography,
floatfix]{revtex4-1}

\usepackage{graphicx}
\usepackage{dcolumn}
\usepackage{bm}
\usepackage{siunitx}
\usepackage{url}

\usepackage{hyperref}
\begin{document}
\preprint{RRR1028}

\title{{\it Colloquium:} Advances in automation of quantum dot devices control}

\author{Justyna P. Zwolak}
\email{jpzwolak@nist.gov}
\affiliation{National Institute of Standards and Technology, Gaithersburg, Maryland 20899\text{,} USA}

\author{Jacob M. Taylor}
\affiliation{Joint Quantum Institute, National Institute of Standards and Technology, Gaithersburg, Maryland 20899\text{,} USA}
\affiliation{Joint Center for Quantum Information and Computer Science,
University of Maryland, College Park, Maryland 20742\text{,} USA}

\date{\today}
\begin{abstract}
Arrays of quantum dots (QDs) are a promising candidate system to realize scalable, coupled qubit systems and serve as a fundamental building block for quantum computers.
In such semiconductor quantum systems, devices now have tens of individual electrostatic and dynamical voltages that must be carefully set to localize the system into the single-electron regime and to realize good qubit operational performance. 
The mapping of requisite QD locations and charges to gate voltages presents a challenging classical control problem. 
With an increasing number of QD qubits, the relevant parameter space grows sufficiently to make heuristic control unfeasible.
In recent years, there has been considerable effort to automate device control that combines script-based algorithms with machine learning (ML) techniques.
In this Colloquium, a comprehensive overview of the recent progress in the automation of QD device control is presented, with a particular emphasis on silicon- and GaAs-based QDs formed in two-dimensional electron gases. 
Combining physics-based modeling with modern numerical optimization and ML has proven effective in yielding efficient, scalable control. 
Further integration of theoretical, computational, and experimental efforts with computer science and ML holds vast potential in advancing semiconductor and other platforms for quantum computing.
\end{abstract}

\maketitle
\tableofcontents{}

\section{Quantum Dot Devices: History and Overview}\label{sec:overview}
The electron is the simplest natural quantum bit. 
It has a spin of $1/2$ and comes with an electric charge, enabling trapping and moving of the particle, much like an ion. 
Furthermore, Kramers's theorem provides an assurance that this simple structure is maintained even in the complex environment of a condensed matter system, specifically the spin-$1/2$ property. 
This means that one can leverage the vast array of condensed matter and solid-state research and the corresponding industry that has developed around the control of electrons, particularly in the case of semiconductors.
Quantum dots (QDs) are one such approach for isolating and controlling single electrons using semiconductor physics \cite{Wiel02-DQD, Loss98-QCD, Hanson07-SQD, Zwanenburg13-SQE}.

\begin{figure*}
\includegraphics{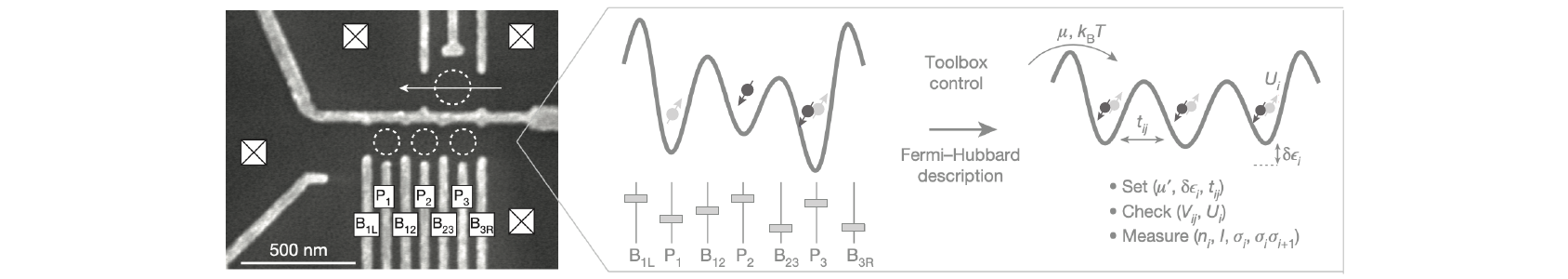}
\caption{Left image: electron micrograph of a gate-defined triple-QD array in GaAs. 
The bottom three dashed circles indicate the qubit QD array, while the single dashed circle with an arrow indicates the sensing QD channel.
Right panel: the goal of tuning QD devices, which is setting a desired Hamiltonian through the efficient control of voltages applied to all gates to enable fine control of QD qubits.
The final form maps to variants of the Hubbard model, where the parameters of detuning, charge interaction, and tunneling are all controlled by gate voltages.
Adapted from \citet{Hensgens17-FHQ}. 
}
\label{fig:intro-fig}
\end{figure*}

Interest in QDs goes well beyond quantum computing, from direct current (dc) standard to low-power logic to thermometry at low temperatures \cite{Likharev99-SDA}.
Working at the single-electron level with both orbital and spin coherence represents the edge of the possible in semiconductors, and thus quantum computing with semiconductor QDs has proven to be a fruitful domain to develop new techniques and uses of isolated electronic systems while maintaining their promise as a pathway to quantum computing \cite{Watson18-TQP, Xue21-CCC, Zwerver22-QMA, Noiri22-FUG, Philips22-UCS, Weinstein22-ULS, Takeda22-QES, Madzik22-PTT}.

The difficulty of working with such systems (where fabrication tolerances are tight, impurities are troublesome, and material considerations are vast) has meant that only a few groups around the world have succeeded in advancing the limit of QD performance, as recently demonstrated through the use of high-fidelity ($>0.99$) two-qubit gates \cite{Noiri22-STL, Xue22-QLS, Mills22-TSP}.
Many of the fabrication and materials challenges have been surmounted with engineering and by a practice of sharing results and even materials and samples throughout the academic community. 
One of the critical remaining challenges is reducing the difficulty of arranging the depletion and accumulation of electrons in a nanoscale semiconductor system such that the right number of isolated islands (QDs) with the right number of electrons (charge state) in the right connection network (topology) is realized every time the system is cooled down and initiated.

This task is unexpectedly difficult. 
A typical gate-based semiconductor QD uses between four and six local depletion and/or accumulation gates to form a single island; see Fig.~\ref{fig:intro-fig}. 
If one seeks to build an array of such devices in close proximity (such that individual electrons can coherently tunnel between islands, as is necessary for most quantum computing implementations) the number of gates scales as some polynomial in at least the square root of the number of QDs, and typically scales linearly \cite{Veldhorst17-SCA}.

To reach a stable few-electron configuration, early experiments set the input voltages heuristically, relying on experimental intuition and informed guesses. 
However, such an approach does not scale well with growing array sizes and is susceptible to random errors.
Moreover, since the parameter space exploration strongly relies on intuition and experience, this tuning may result in only an acceptable rather than an optimal configuration. 
Finally, the size of the relevant space of parameters that need to be adjusted with an increasing number of QD qubits makes heuristic control even more challenging.
As a result, there has been significant interest in exploring the potential of machine learning (ML) to solve the autotuning problem. 

We now consider the specific challenges of bringing a device into the desired operating regime. 
An essential ingredient on any tuning procedure is determining the state of the device (topology, number of charges, parameters such as tunneling rates, etc.). 
However, measuring these devices is nontrivial. 
In general, the field has progressed by exploratory measurement of electrical currents through nearby circuits\cite{Field93-MCB, Elzerman04-SRE}. 
These techniques rely upon the variable conductance of the underlying semiconductor, which in turn depends upon the local electric field produced by nearby electrical gates or by electrons that are trapped in the QDs themselves. 
The probe systems must be nearby, as screening of the electric field from nearby metallic regions reduces long-range coupling and corresponding changes in conductance become hard to measure.

These current-flow approaches commonly use an auxiliary QD or quantum point contact to act as a sensor of the local electric field and are sufficiently sensitive to measure the change in the local potential landscape to the addition or subtraction of a single nearby charge \cite{Simmons07-QCS,Lai11-PSH,Petit20-UQL,Yang20-OQK,Blumoff22-FPM,Mills22-TSP}.
Their ability to work independently of coupling of the QDs in the main device provides substantial engineering simplicity in creating and tuning QD devices.

In addition to charge-based measurements, there are other measurement techniques. 
The first was more common in the earlier era of the field: directly measuring the current flow through the QD of interest \cite{Hendrickx20-FTL, Maurand16-CSS}. 
This relies upon finding degeneracies between all the relevant charge configurations in the QDs such that electrons can freely flow from a source to a drain. 
However, these are not as useful for tuning to the single-electron regime, as the current flow through the devices typically gets small as one approaches that regime, and tuning to these degeneracy points becomes increasingly challenging as the number of devices in series increases. 
Nonetheless, some ML-based autotuning techniques operate directly on current-flow data \cite{vanEsbroeck20-FTQ}.

Furthermore, using charge sensors as previously described is necessary but not sufficient for a complete tune-up, as fine-tuning of devices, and the eventual characterization of their quantum gates, requires single-shot readout of individual charge configurations at the end of a sequence of pulses. 
Thus, more complex sensing routines, including time-sequenced high-frequency gate control and rf or microwave reflectometry, must also be integrated and calibrated.
Developing these capabilities and leveraging them earlier in the tuning process is generally beneficial, as more rapid data acquisition leads to more rapid tuning: typically, automation is limited by data from the experiment, not by classical computing costs.

Finally, there are some new directions in direct integration of circuit quantum electrodynamic techniques for measurement.
A typical example eschews the use of a separate charge sensor, instead connecting one of the gates directly to a superconducting resonator. 
The local response of the QD system to an applied microwave or radio frequency field then causes the field in the cavity to have a phase shift; in essence, the QD system acts as a variable capacitor in a resonant circuit \cite{Mi17-SCS, Crippa19-GRR, Zheng19-RGS, Liu21-RFR, Burkard20-SSH}. 
Thus, even though no current flows through the QDs, one can directly measure the local properties about the response of energy levels to the applied fields. 
This has shown promise outside the ML space for fine-tuning, as it enables one to estimate other properties, such as valley degeneracy in silicon devices \cite{Mi17-HVS}. 
However, this gate-based readout can lead to difficulties in seeing the global charge state, and thus may cause overall features necessary to get to a particular topology to be missed. 
On the other hand, the approach is likely to be effective for fine-tuning once the topology and number of electrons is configured.
We note that there are other techniques for extracting this information as well, such as fast pulses over anticrossings, the use of large bias currents, or even elevation of the temperature of the device.

In this Colloquium, we endeavor to both set up the challenges of tuning QD systems and describe progress toward their automation. 
We begin with an overview of QD qubits, followed by a discussion of the Hubbard model representation of the QD system and a high-level overview of the QD tuning subprocesses that take a system from room temperature to quantum operation.
Several key metaproblems related to scalability and tuning in the presence of imperfections are also discussed.
The concept and methods of tuning QD devices are presented in Sec.~\ref{sec:qd-today}.
The overview and discussion of present-day advances in tuning automation are presented in Sec.~\ref{sec:towards-autom}.
Section~\ref{sec:outlook} concludes the Colloquium with a summary and outlook, including some questions related to scaling up and ``on-chip'' implementations of the autotuning systems.
Closing remarks are presented in Sec.~\ref{sec:concl}.

To date the techniques described here are largely a replacement of laboratory heuristics with a combination of workflow-based automation and key ML subsystems. 
However, the rapid advances in ML suggest that future work will more directly address the global on-line optimization problem that tuning presents and also find new means of replacing larger parts of today’s workflow with faster and more accurate methods. 
We await these future directions with optimism.

\section{Quantum Dot Devices Today}\label{sec:qd-today}
QD systems can be used to form qubits, where the spin, valley, internal electronic state, or the charge state of the electrons in one or several QDs is used to store quantum information. 
This information is then manipulated by control of global and local magnetic and electric fields, typically in a narrow range of the charge parameter space.
Common examples of QD-based qubits include using the electron spin in a single dot (up and down representing a qubit, leveraging Kramers’s theorem to ensure that such states exist), using a pair of electron spins in adjacent QDs in which the parity of the two spins stores the quantum information, or using the exchange symmetries of three electron spins spread over three QDs in a so-called exchange-only qubit.
In spin qubits, the different individual spin configurations of the same charge configuration define the logical states, with up and down defined relative to an external magnetic field \cite{Loss98-QCD, DiVincenzo00-UQE, Vandersypen17-ISQ, Burkard21-SSQ}.
For composite qubits, the logical states of the qubit are defined by symmetry operations on spin configurations resulting from having two, three, or even four electrons distributed among two, three, or four QDs \cite{Shi12-HSQ, Cao16-THQ, Petta05-CMQ, DiVincenzo00-UQE, Russ17-TES, Wu14-TCQ, Sala17-ESS, Russ18-QES}.
In principle, charge qubits, where the logical states of the system are defined by different charge occupations of adjacent tunnel coupled QDs \cite{Kim15-MDC, Gorman05-CQO, Petersson10-QCC}, are also a path toward qubits.
In all cases, the critical task of getting a QD system into a regime in which qubits can be realized has mostly common features, with only minor changes at the final stage of tuning to distinguish between the different qubit types.

In many respects, this Colloquium is focused on developing the tools to get one to a regime in which the previously mentioned qubit types can be reliably realized. 
From a systems engineering perspective, a useful intermediate representation of the system is the Hubbard model, in which individual QDs labeled with index $i$ are allowed to have zero, one, or two electrons on them at a time \cite{Loss98-QCD, Burkard99-CQG, Yang11-GHM, DasSarma11-HDQ}. 
It is convenient to define a site occupation number $n_i = \sum_\sigma c_{i,\sigma}^\dag c_{i,\sigma}$, with the fermionic creation and annihilation operators $c_{i,\sigma}^\dag$ and $c_{i,\sigma}$, respectively, and $\sigma = \pm 1/2$ being the spin, to represent the number of electrons on site $i$.
Therefore, we have
\begin{equation}\label{eq:hubbard}
\begin{split}
    H = &-\sum_{ij} t_{ij} (c_{i,\sigma}^\dag c_{j,\sigma} + {\rm H.c.}) \\
    &+ \sum_{ij} \tfrac{1}{2} (n_i - \epsilon_i) U_{ij} (n_j - \epsilon_j) + H_{\rm B}
\end{split}
\end{equation}
where the first sum is over QDs that are tunnel coupled with tunneling energy $t_{ij}$ to each other. 
The second term of Eq.~(\ref{eq:hubbard}) incorporates $i$th site occupation $n_i$, charge offsets due to gate voltages $\propto \epsilon_{i(j)}$, and the inverse capacitance matrix $\sim U_{ij}$ that determines the lowest energy charge configuration. 
We note that the energy relationship of capacitance $(1/2) Q C^{-1} Q = (1/2) V C V$ is exactly that captured by $U_{ij} \approx e^2 C^{-1}$, where $e$ is the single-electron charge. 
As this matrix is symmetric, the sum runs over all $i$ and $j$, although it can be redefined as a sum over $i \geq j$.
Later we use this relationship to define virtual gates.
The last term in Eq.~(\ref{eq:hubbard}) $H_B$ is the magnetic field Hamiltonian for the spins only and typically includes both a uniform component and a local (gradient) component. 
As mentioned, beyond the charge state QDs have internal electronic, spin, and valley levels, that become particularly important for qubit operation. 
Thus, in general, a multiband approach to the Hubbard model should be used for real experiments.
However, since these complications do not affect most of the tune-up process except at the final (fine-tuning) stage, we do not include them here, as they rarely impact the charge-tuning process.
Automated procedures (described later) can help directly extract the functional dependence of the tunneling ($t_{ij}$) and capacitance ($U_{ij}$) terms on the device parameters, such as gate voltages, as well as further voltage dependencies in $\epsilon_{i(j)}$. 

\begin{figure}[t]
\includegraphics[width=\linewidth]{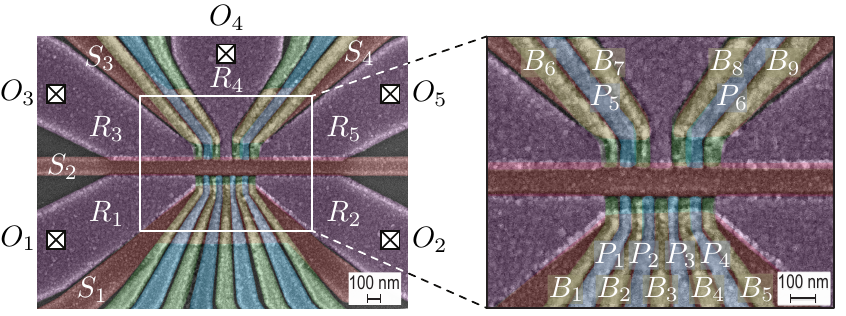}
\caption{False-color SEM of a Si/SiGe quadruple-QD device. 
This gate architecture is used as an example when we discuss the autotuning process.
All gates and voltage sources are labeled with standard names: screening ($S$), reservoir ($R$), plunger ($P$), barrier ($B$), and Ohmic contact ($O$).
The upper channel of this device with two QDs and a central Ohmic contact enables each to be used as a charge sensor for the lower channel QDs. 
Adapted from \citet{McJunkin21-PhD}.
}
\label{fig:device-scan}
\end{figure}

To get to Hubbard abstraction, we now consider a typical gate-defined QD device in Si/SiGe that is shown in Fig.~\ref{fig:device-scan}.
The different metal gate electrodes are intended to serve different functions in forming and controlling the QD qubits by depleting or accumulating electrons in the semiconductor underneath the gates. 
The so-called finger gates, named for their resemblance to the fingers on a hand, control the tunnel coupling with the reservoirs $\Gamma_i$ (barrier gates $B_{i}$, where $i=1,5,\dots,9$, in Fig.~\ref{fig:device-scan}) and the interdot tunnel couplings $t_{ij}$ (barrier gates $B_{i}$, where $i=2,3,4$, in Fig.~\ref{fig:device-scan}) and are used to set the electron number in each QD by changing the local electrostatic potential (plunger gates $P_{i}$, where $i=1,\dots,6$, in Fig.~\ref{fig:device-scan}).
However, in practice each gate influences the other parameters as well (due to the capacitive crosstalk), as we later discuss.
In a typical setting, one of the QD regions is separated out from the others (top area of Fig.~\ref{fig:device-scan}) and is used as a local probe of the electrostatic environment, thus measuring changes in the number of nearby charges.
This measurement system allows for assessment of the rest of the device and provides a helpful abstraction from currents and conductances.

\begin{figure*}
\includegraphics[width=\linewidth]{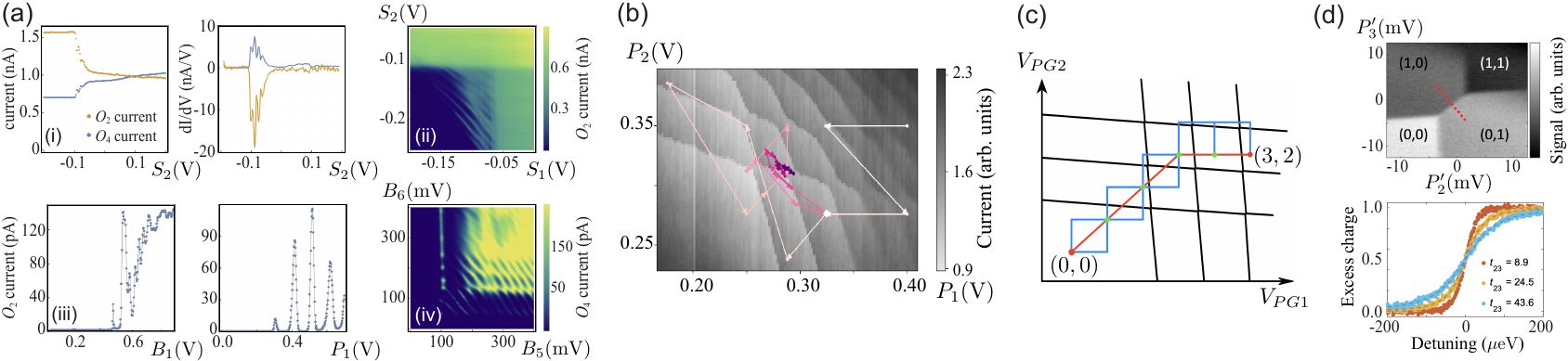}
\caption{Autotuning flow.
(a) Sample plots used in the bootstrapping phase described in Sec.~\ref{ssec:boot}; see Fig.~\ref{fig:device-scan} for a SEM of a device showing the gate nomenclature.
(i) Pinch-off measurement of the QD current $O_i$, where $i=2,4$ (left panel) and the numerical derivative of the measurement (right panel) for the central screening gate $S_2$ used to divide the 2DEG into two half planes. 
Current through both the qubit and sensor side of the QD system is measured as the gate is swept to negative voltages.
(ii) Example of a triangle plot of the qubit QD channel. 
The large rectangular regions to the right and at the top represent current flow underneath two screening gates.
(iii) A series of pinch-off curves for finger gates showing Coulomb blockade oscillations.
(iv) The so-called wall-wall plot in the space of the two barriers
defining the sensor. 
From \citet{McJunkin21-PhD}.
(b) Visualization of a state tuning run in the space of two plunger gates, with the arrows and the color intensity indicating the progress of the autotuner.
From \citet{Zwolak20-AQD}.
(c) Sketch of a possible path (red line) connecting the $(0,0)$ and the $(3,2)$ charge state.
From \citet{Durrer19-ATQ}.
(d) Top panel: charge stability diagram in the space of virtual gates with a red dotted line indicating the interdot detuning axis. 
Bottom panel: excess charge extracted from a fit to the sensing QD signal as a function of detuning along the dotted line for three different $t_{23}$. 
From \citet{Hsiao20-EOT}.
}
\label{fig:tuning-phases}
\end{figure*}

The tuning process is an essential, albeit repetitive, step for initialization of QD-based qubits.
The process of tuning an unknown QD device can be divided into a sequence of distinct phases, \emph{bootstrapping}, \emph{coarse tuning}, \emph{charge state tuning}, and \emph{fine tuning}, as shown in Fig.~\ref{fig:tuning-phases}. 
Depending on the setup, an additional \emph{establishing controllability} phase may be carried out after the charge-state tuning phase to enable targeted gate control.
We note that, given the dependence of the transformation to control space on the actual charge states, the controllability step may need to be executed more than once throughout the tuning process.
There have been numerous efforts around automating each of those phases, as we discuss in Sec.~\ref{sec:towards-autom}. 
Here we present a high-level overview of what each of those phases encompasses and the desired output for each. 

\begin{description}
    \item[\it{Bootstrapping}]---Bootstrapping (or initialization) is a pretuning process that involves cooling the device down, making local sensing systems operational, and bringing the main device regime into an appropriate parameter range for taking data (i.e., measurement setup). 
    This might include bringing charge sensors into operation and preparing a 1D channel where QDs are going to be formed in the two-dimensional electron gas (2DEG).
    Bootstrapping can also build a ``sandbox'' of acceptable parameter variations by checking the finger gates for response and pinch-off voltages and determining the safety ranges for coarse tuning. 
    
    \item[\it{Coarse tuning}]---Coarse tuning is a process of finding a range of gate voltages where the device is in a particular global configuration corresponding to a set of charge islands (QDs) with an understanding of their connectivity, i.e., which island is tunnel coupled to which. 
    In simple devices, this corresponds to an estimate such as that of a no-QD, single-QD, or double-QD regime.
    As such, it can be thought of as defining the device topology in the state space.
    At the end of this phase, the QD device should be in a stable global configuration of known topology. 
    This topology is the output of the coarse-tuning process.
    
    \item[\it{Establishing controllability}]---Ideally, changing voltages on a single gate affects only the parameter it is designed to control (such as the electrochemical potential of a specific QD or the tunnel barrier between two adjacent QDs).
    However, the capacitive crosstalk between the various gate electrodes (i.e., a situation in which a signal on one gate causes a smaller version of the same signal on an adjacent gate because of the capacitance between them) makes it difficult (if not impossible) to vary only a single parameter without affecting the others.
    One way to compensate for the capacitive crosstalk is to implement so-called virtual gates, that is linear combinations of multiple QD gate voltages chosen to address only a single electrochemical potential or tunnel barrier.
    Virtual gates are the output of the establishing controllability phase.
    
    \item[\it{Charge state tuning}]---While at the end of the coarse-tuning phase the device is in a definitive global state (i.e., single or double QD), the number of charges on the QDs is at this point unknown.
    Since each QD qubit type requires a specific charge arrangement, the goal of charge-state tuning is to bring the QD device to a specific charge configuration, i.e., the specified number of charges (typically one to three) on the specified islands.
    
    \item[\it{Fine tuning}]---Once the system is in the desired topology and electron number, the system is ready for preparing it for use as a quantum computing or quantum simulation device.
    This stage of tuning (fine tuning, as we call it here) requires tremendous precision and expertise in order to go from something that has roughly the right Hamiltonian and charge configuration to a fully controlled multiqubit system. 
    For example, interdot tunnel couplings and dot-lead tunnel couplings have to be precisely tuned and calibrated the electron response to applied voltages at the quantum mechanical level needs to be mapped out, the valley degeneracies and magnetic gradients must be understood, translations of microwave signals to applied pulses on the electrons have to be determined, etc.
    This fine-tuning work is well beyond the scope of this Colloquium, although many of these elements bear a strong resemblance to similar challenges currently being addressed in superconducting and ion-trap-based quantum computing designs, such as those used by \citet{Klimov20-SOQ} and \citet{Kelly18-QDG}, and we may be able to leverage advances in those fields with little additional QD-specific modifications.
    Thus, in what follows we focus on the physics elements of these calibrations only, leaving the quantum computing elements (including qubit initialization, manipulation, and readout) to future work.
\end{description}

Beyond these elements, there are several key metaproblems that should be addressed in future work. 
The most prominent is related to the ability to scale QD devices. 
Specifically, to what extent can we take a tuning system for one of the previously mentioned tasks and while holding a portion of the QD device tuned, tune up a different portion of the device? 
For example, in a long array of QDs, one might first tune up a nearby charge sensor.
One might then tune a double QD. 
What if one seeks to add another QD to the chain? 
This inductive tuning problem has been shown to work experimentally \citep{Volk19-LQR}, but theoretically there are substantial open questions about the ability to rapidly and reliably tune a large 1D array. 
As an example of why this is challenging, it is often the case in higher-dimensional arrays for there to be charge configurations that are lower in energy than the current configuration but that the device does not reach due to metastability; that is, there is no easy way for an electron at, say, the edge of an array of QDs to make its way to the relevant QD near the center of the device to let the system relax to its true charge ground state. 
Sometimes this problem is called latching, as the presence of an electron in a metastable configuration can prevent other electrons from tunneling to get the entire system into its ground state \cite{Volk19-LQR}. 
This metastability in turn can lead to hysteresis when tuning QD devices, as the history of tuning leads to being stuck in different metastable configurations.

A second key metaproblem is how such tuning systems behave in the presence of real-world imperfections. 
From the efficiency standpoint, it would be beneficial to determine whether the QD devices in a wafer are suitable for tuning prior to the wiring and cooldown.
This is particularly important for high-throughput production, where individual wafers may contain hundreds or even thousands of devices. 

Tuning up a multithousand-QD device can be accomplished only if the systems can be solved semi-independently (using recursive or inductive solutions).
It must be done repeatably, in the presence of expected noise, and must respond appropriately to unexpected events and quickly enough to maintain its final state before going ``out of tune.'' 
A large-scale system achieving these tasks is the main target for future research in this direction.

As a concluding note, there are a variety of ways to trap individual electrons in a semiconductor system. 
In this Colloquium, we focus on devices where the electric field plays a key role in accumulating or depleting regions of a semiconductor, all the way down to the single-electron level. 
This encompasses a large variety of different technologies and materials. 
In many cases, these systems have qualitative similarities. 
However, there are both large and subtle differences that can make ``porting'' a tuning method from one material (such as GaAs) to another (such as silicon-based QD) more complex or even likely to fail. 
For example, in fine-tuning silicon devices have valley degeneracy and a large effective electron mass, leading to substantially different behavior in the $0.1~\si{\milli e\volt}$ energy scales of orbital and related physics. 
As another example, devices with electrons close to oxide interfaces are known to have a large probability of so-called spurious QDs, in which an impurity or defect causes the formation of a charge trap that can impact the controllability and measurability of the QDs. 
Finally, differences in noise, dielectric properties, and gate geometries mean that each application of a tuning method must generally be targeted toward both a specific material and a specific approach to QD formation.
In what follows, we focus on silicon- and GaAs-based QDs formed in 2DEGs but note that many of these techniques can be generalized, with the previous proviso.

\section{Toward Full Automation of Device Control}
\label{sec:towards-autom}
There have been several attempts at automation of the various steps of the tuning process over the past decade.
The initial script-based approaches rely on intuitive and relatively easy to implement conventional algorithms.
They typically involve a combination of techniques from regression analysis, pattern matching, and quantum control theory.
However, the defects and variations in the local composition of the heterostructure, as well as fabrication variances that disorder the background potential energy, make semiautomated control challenging.
The random disorder that must be compensated for by different gate voltages when defining QDs further impedes the script-based tuning process.
None of the proposed approaches to date fully eliminate the need for human intervention, as the decision on how to adjust gate voltages, based on the automatically obtained quantitative QD parameters and qualitative output, remains heuristic.
These approaches are typically calibrated to a particular device architecture (or even a specific device), are susceptible to noise, and do not transfer well to other devices~\cite{Baart16-CAT}.

More recently researchers began to take advantage of the tools provided by the field of artificial intelligence and, more specifically, supervised and unsupervised ML.
ML is a branch of computer science that focuses on the use of data and various specially designed algorithms to imitate the way that humans learn.
Throughout training, a ML algorithm learns or, in some applications, discovers patterns in data without being explicitly programmed about the characteristic features of those patterns. 
Thus, if one is provided with proper training data, ML-enhanced methods have the flexibility of being applicable to various devices without any adjustments or retraining.
Since they learn from the data, ML algorithms are more adjustable to new devices than script-based methods.
However, ML models typically require large labeled datasets for training and often lack information on the reliability of the ML prediction~\cite{Zwolak18-QLD, Darulova20-EDM}.

While the problem of estimating and tuning systems is generically challenging, QD-based devices do benefit from having a variety of distinct problems and means of estimation that are sufficiently mature to enable applications of specific ML techniques to various subproblems without requiring a holistic solution. 
However, there may be some benefits to considering a holistic approach, particularly in low information settings such as those considered by \citet{Moon20-ATQ}.

In Sec.~\ref{ssec:boot}--\ref{ssec:fine}, we describe advances in QD autotuning while focusing on the five main phases of the tuning process discussed in Sec.~\ref{sec:qd-today}.
In general, all techniques used for automation follow one of the two main schools of thought: computer-supported, script-based gate control or ML-driven methods.

\subsection{Bootstrapping and sandboxing quantum dot device}
\label{ssec:boot}
The mobility of electrons (or holes) as well as many-body interactions between an isolated spin trapped in a QD and the sea of nearby spins in the 2DEG are often strongly dependent on a variety of parameters.  
Of particular relevance for observing quantum coherent behavior is the overall temperature of the device, as charge-phonon coupling prevents coherent behavior at moderate temperatures, and the necessary effects, such as resolution of excited electronic states in an individual QD, become impossible for temperatures well above the energy regime of interest.
In general, the weaker the interaction, the colder the 2DEG has to be in order to observe it.
For example, the smallest energy splitting in silicon QDs, the valley splitting, is typically 10 -- 300~\si{\mu e\volt} for Si/Si$_x$Ge$_{1-x}$~\cite{Simmons10-PLD, Borselli11-MVS} and about 300 -- 800~\si{\mu e\volt} for silicon metal-oxide semiconductors (SiMOSs) \cite{Yang13-MVS}, which is small compared to room temperature (1~\si{\kelvin} is 87~\si{\mu e\volt}).
Cooling down the device is the first step of the bootstrapping phase. 
The usual technique for cooling 2DEG samples to the temperature required for an experiment is to use a dilution refrigerator.
Once the sample is sufficiently cooled, pretuning and functionality testing can begin.

At a basic physical level, the underlying semiconductor chip will have a variety of reasonable ranges of voltages that can be applied, which is often determined ahead of time in a probe state or, in the case of an industrial lab, can be engineered. 
This allowed range is designed to prevent undesired effects, including electrostatic breakdown, leakage of current across the barriers, and other similar effects. 
Thus, a basic level of device protection must be assured by providing limits that the control systems will respect.
However, there is a much narrower range of desired operation than that restricted by pure physical damage. 
Determining this range is one of the goals of bootstrapping. 
Depending on the device type, there are two additional goals: (i) assuring that a 1D transport channel is formed in the 2DEG (not applicable to nanowire QDs) and (ii) properly calibrating the charge sensor (not applicable to devices that do not use charge sensors, although this will typically have to be replaced with other constraints for either electrodynamic or current-based readout).

As an example, we consider silicon QDs in devices with an overlapping gate architecture, which require both goals (i) and (ii). 
In these devices, the active region for forming QDs is provided by a 2DEG formed at the interface between layered semiconductor structures (heterostructures), where the free electrons are confined to a planar region so thin that they behave as if they are truly two dimensional.
An example of a false-color SEM of a sample device with the overlapping gates architecture is shown in Fig.~\ref{fig:device-scan}.
The functionality testing of such devices begins with an establishment of the so-called global turn-on, that is, determining the voltage level at which a 2DEG is uniformly accumulated underneath the gates \cite{McJunkin21-PhD}.
This is achieved by synchronously increasing the voltage applied to all gates (initially set to $0$~\si{\volt}) while simultaneously measuring the current through appropriate pairs of metal Ohmic contacts.
The level at which sufficient conductance is observed indicates the turn-on voltage.

The next step for this device is to assure that two 1D channels are formed within the 2DEG: one where the qubit QDs will be formed and the other to host the charge sensing QDs. 
This is achieved by properly adjusting the screening ($S_i$, where $i=1,\dots,4$, in Fig.~\ref{fig:device-scan}) and reservoir ($R_i$, where $i=1,\dots,5$) gates .
The 2DEG first has to be divided into two separate planes (current paths) by depleting electrons under the central screening gate [$S_2$ in Fig.~\ref{fig:device-scan}; see also the pinch-off curves in Fig.~\ref{fig:tuning-phases}(a)(i)].
To ensure sufficient accumulation and uniformity of the 2DEG, it might be desirable to readjust the turn-on for each side separately.
Since the current through QDs has to be controlled at the single-electron level, it is important to verify that reservoir gates are not overaccumulated (or underaccumulated).
This is done by testing the pinch-off of each reservoir, that is, finding the voltage level that cuts off the current through the device.

The remaining screening gates then need to be calibrated to form two narrow conduction channels on each side of the device.
By pinching off an appropriate set of screening gates [such as $S_1$ and $S_2$ for the qubit channel in Fig.~\ref{fig:device-scan}; see also Fig.~\ref{fig:tuning-phases}(a)(ii)] and then setting the voltages on them significantly below the pinch-off, it is guaranteed that the 2DEG under those gates is depleted.
Thus, if a current flow is observed when voltage is simultaneously increased at all relevant finger gates (i.e., gates $B_i$ for $i=1,\dots,5$ and $P_j$ for $j=1,\dots,4$ for the qubit channel), it can be safely assumed (for the overlapping gate design) that the 2DEG is accumulated only in the narrow passage between the screening gates.

Once the 1D channel is formed, quality assessment and characterization of the finger gates can begin.
This process serves two purposes: (i) if a given gate does not pinch off the current, it might be an indication that the gate is defective or broken and thus the devices should not be further calibrated, and (ii) the pinch-off values for each individual gate can be used as a guide when setting up the voltages needed to form QDs.
Two pinch-off curves showing Coulomb blockade oscillations in properly working finger gates are depicted in Fig.~\ref{fig:tuning-phases}(a)(iii). 
As mentioned, voltages applied to the finger gates generally have to compensate for the disordered potential landscape resulting from various fabrication defects.
Since the disorder is largely random and varies between devices, the optimal voltage for each gate not only has to be determined for each new device but also might be necessary between cooldowns of the same device.
During the functionality testing step, all gates are checked and the optimal voltages are determined.

Typically, all of the bootstrapping steps are performed through a sequential analysis of 1D measurements of transport through the device as a function of gate voltages.
As such, this process is well suited for automation.
One way to achieve this goal in the context of gate-defined QD devices is to combine physics-informed fitting and thresholds to inform the selection or relative voltage ranges as well as consecutive adjustments \cite{McJunkin21-PhD, Baart16-CAT}.
An approach to characterizing gates involving fitting and binary classification has also been proposed \cite{Darulova19-ATQ}.
In this approach, a set of parameters defining a hyperbolic-tangent-based fit to the 1D measurements was extracted and used to define, among other things, the pinch-off, transition, and saturation regions for each gate.
A subset of these parameters was then used as a proxy for measurement quality (classified as good or bad) as well as to define the voltage range for subsequent coarse tuning.
However, this approach is applicable mainly where not much variability in the pinch-off curves between the gates is expected. 
This is commonly seen in GaAs devices, while it is observed much less often in silicon devices. 
An automated analysis of the pinch-off curves for a device with an overlapping gate architecture becomes significantly more challenging as Coulomb blockade oscillations become present in the transport measurement. 
Specifically, depending on the finger gates' location (such as the proximity to the reservoir) and fabrication discrepancies, the pinch-off curves for gates on a single device can exhibit significantly different behavior, as depicted in Fig.~\ref{fig:tuning-phases}(a)(iii), with the left panel showing only some Coulomb blockade oscillation while the oscillations in the right panel are much more rapid (likely due to a large level arm between the gate and the dot).
A simple threshold based on a tanh-like fit [as implemented by \citet{Darulova19-ATQ}] might be sufficient to extract the pinch-off, transition, and saturation regions from the left curve; however, it would be insufficient for the right curve. 
For the latter, a more advanced detection of Coulomb blockade oscillations would be beneficial.
Thus, while these measurements can still be used to determine the pinch-off, a more nuanced analysis of the oscillations might be desirable to determine the transition width of the pinch-off curve.

The final step of bootstrapping is calibration of the charge sensor, that is, finding a set of voltages applied to the sensing QD gates that maximizes sensitivity of the conductance through the QD to changes in the local electrostatic potential.
This is typically done through an analysis of a 2D plot in the space of the two barriers defining the sensor starting with both barrier gates set slightly below their respective pinch-off values and sweeping until saturation (as determined by the pinch-off width) \cite{Baart16-CAT, Botzem18-TSD, McJunkin21-PhD}.
The resulting plot[the so-called wall-wall plot, see Fig.~\ref{fig:tuning-phases}(a)(iv)] should reveal Coulomb blockade oscillations indicating the voltage range where the tunneling through the QD is finite, as indicated by the characteristic diagonal lines.
There are numerous edge detection techniques that can be deployed to identify the position and extract the locations of those lines.
Alternatively, a series of 1D sweeps of one of the barrier gates (from below the pinch-off to saturation) while keeping the other barrier fixed slightly above its pinch-off level should also reveal the oscillation pattern.
Either method should produce voltages for setting the barriers for the charge sensor plunger sweep.
The final 1D measurement from the plunger pinch-off until saturation should reveal strong Coulomb blockade oscillations, thus indicating that this tuning is sufficient for operation as a charge sensor.
The best charge sensitivity is achieved by choosing a voltage combination where the slope of the Coulomb peak is steepest.

\begin{figure*}[t]
\includegraphics[width=\linewidth]{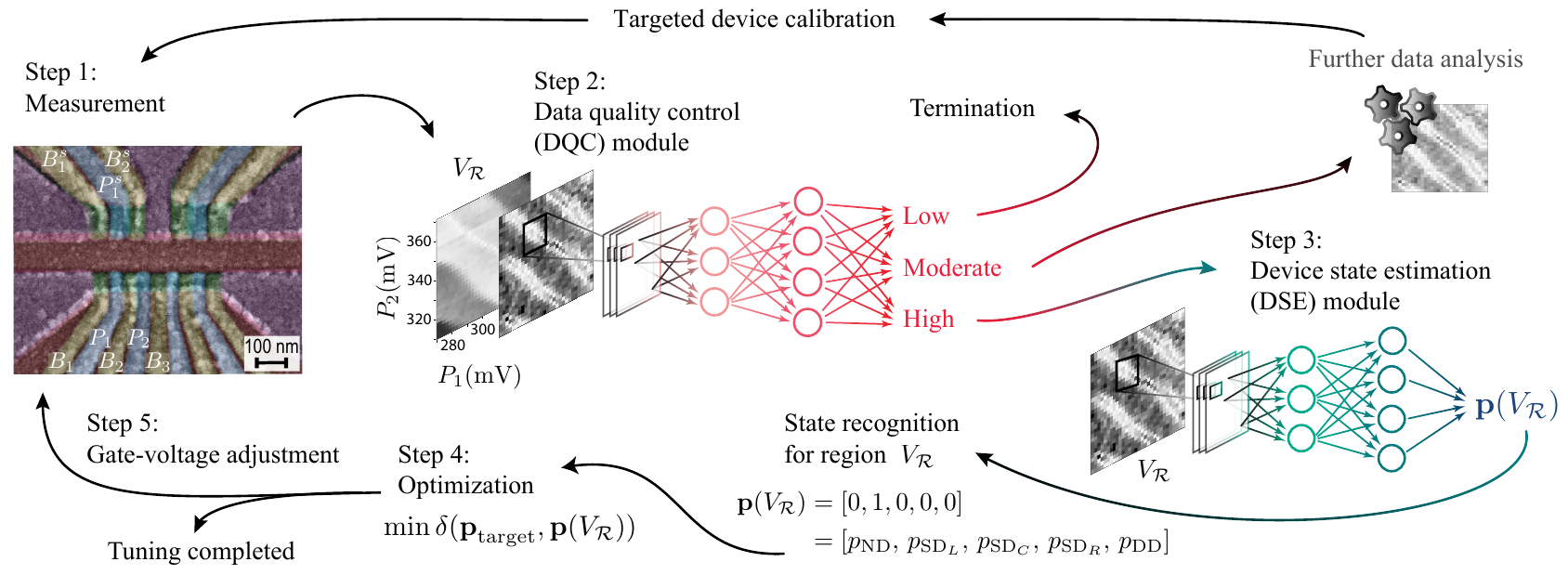}
\caption{Visualization of the coarse-tuning loop. 
The data processing step introduced by \citet{Zwolak20-AQD} is replaced by the data quality control module (DQC; step 2). The DQC is followed by state assessment and optimization for a high-quality scan; additional data analysis and targeted device recalibration for a moderate-quality scan; or the autotuning loop termination for a low-quality class. 
Adapted from \citet{Zwolak20-AQD} and \citet{Ziegler22-TRA}.
}
\label{fig:ml-coarse}
\end{figure*}

\subsection{Coarse tuning: Setting device topology}
\label{ssec:coarse}
Once the two 1D channels are formed and the charge sensor is ready, the next step in the tuning process is setting the QD device up in a stable global configuration of known topology in the state space with a known number of charge islands. 
The coarse-tuning phase is a process of finding a range of gate voltages for all the finger gates on the qubit side of the QD device that set the device in a particular global configuration.
For simple devices, this corresponds to bringing the device into a single- or double-QD state, preferably in the few-electron regime.

\vspace{-4pt}
\subsubsection{Topology of single-QD devices}
\label{sssec:sd_coarse}
The process of tuning to a single-QD state is similar to that of tuning a charge sensor: finding pinch-off levels for all gates defining the QD, followed by calibration of barriers, and then sweeping the plunger gate.
The Coulomb peak oscillations present in the resulting plot correspond to discrete charge transitions through the QD, indicating that a single QD has been formed.
Depending on the QD device type and architecture, tuning to a double-QD state may involve adjusting the plunger gate until the single QD separates into two charge islands or tuning the adjacent set of gates to a single QD followed by fine-tuning of the crosstalk between the two neighboring QDs.
At each step, the state of the device is determined based on visual inspection of the data.
The various possible states manifest themselves as different shapes formed by electron transition lines, sets of parallel lines and honeycombs for the single- and double-QD states, respectively.
Moreover, the orientations of those lines with respect to the measurement direction in the gate voltage space allows determination of the actual location of the charge island as a left, central, or right single QD.  

Both script-based \cite{Lapointe-Major19-ATQ} and ML-enhanced \cite{Czischek21-MNA} automated tuning protocols that combine signal processing and adaptive measurement sequences derived from heuristics have been investigated for SiMOS single-QD devices.
The sequences of measurements implemented in these algorithms are designed to mimic a typical manual tuning process, that is, adjusting voltages on gates defining the QD until transition lines become visible.
The proposed algorithms involve an analysis of a series of images, each capturing a small subregion in the voltage space, for a presence of lines.
If no transitions are detected, both gate voltages are being simultaneously decreased by a fixed amount, creating a diagonal series of measurements until either a transition line is detected or a consecutive adjustment of voltages would surpass a safety restriction.
Once a transition line is detected, the charge-tuning stage is initiated, as discussed in Sec.~\ref{ssec:charge}. 

\vspace{-4pt}
\subsubsection{Topology of double-QD devices}
\label{sssec:dd_coarse}
A first computer-automated approach to tuning double-QD devices was proposed by \citet{Baart16-CAT}, where a GaAs quadruple-QD device was used to demonstrate an algorithm combining various image processing techniques to bring the device to a double-QD state in a single-electron regime.
The expected voltage ranges necessary to observe the Coulomb oscillations were determined by first fitting a tetragon to the area corresponding to large current in 2D barrier-barrier scans and then applying a Gabor filter to identify the center of the specific location where Coulomb peaks were formed.
The resulting gate voltages serve as a starting point to form two independent single QDs.
To form a double QD, a heuristic-based formula accounting for the capacitive coupling of the gates to the QDs was used to determine the correct scan range.
Finally, a pattern matching with a reference template containing expected geometry of the transition lines crossing was used to confirm that a double-QD regime was reached.
However, while in principle only prior knowledge of the gate design and the pinch-off value of the single gate shared by all QD is necessary to initiate this algorithm, in practice the consecutive steps rely on substantial knowledge about the device (such as  a voltage required for the plunger gate to create a singe QD, a suitable scan range for 2D scans, or the expected relative orientation of the transition lines in the crossing template).
This makes the proposed tuning method not easily adaptable to new and unknown QD devices.

ML as a path toward scalable automation of tuning QD devices was suggested for the first time in the context of coarse tuning by \citet{Kalantre17-MLD}.
In the proposed autotuning framework, the laborious tasks of a visual inspection and analysis of 2D measurements by a trained expert is replaced by a ML system [specifically a convolutional neural network (CNN)] trained to quantify the captured state of the device.
The prediction vector returned by the CNN classifier (the ``state vector'') represents the probability of each possible state being present within a given measurement.
For the case of a double-QD system this would be
\begin{equation}\label{eq:prob_vec}
{\rm\bf{p}}(V_\mathcal{R})=[p_{\rm ND},\,p_{{\rm SD}_L},\,p_{{\rm SD}_C},\,p_{{\rm SD}_R},\,p_{\rm DD}],
\end{equation} 
where $V_\mathcal{R}$ denotes the measured scan, ND indicates that no QDs formed, ${\rm SD}_L$, ${\rm SD}_C$, and ${\rm SD}_R$ denote the left, central, and right single-QD states, respectively, and ${\rm DD}$ denotes the double-QD state.

Once properly trained, the ML classifier is integrated with a classical optimization algorithm to control the QD device and navigate the voltage space by minimizing some distance function $\delta(\rm{{\bf p}_{target}},\rm{\bf{p}}(V_\mathcal{R}))$ between the predicted $\rm{\bf{p}}(V_\mathcal{R})$ and target $\rm{{\bf p}_{target}}$ state vectors, as shown in Fig.~\ref{fig:ml-coarse}.
This autotuning framework was originally validated off-line using premeasured experimental scans capturing a large range of gate voltages \cite{Kalantre17-MLD} and then deployed on-line (i.e., \emph{in situ}) to tune a double QD in real time \cite{Zwolak20-AQD}.
However, while the tests showed a lot of promise, with a success rate of $85.7~\%$ over 14 different on-line tuning runs of the same device, the experimental implementation required device-specific data processing to ensure compatibility with the CNN model trained on simplistic noiseless synthetic data.
Still, when data were unusually noisy due to instability of the charge sensor, the data processing was insufficient, resulting in the CNN returning an incorrect state vector and, ultimately, tuning failure.

To prevent failures due to low data quality, \citet{Ziegler22-TRA} recently proposed replacing the data processing step from the original autotuning framework with a ML-driven data quality control (DQC) module.
In the revised framework shown in Fig.~\ref{fig:ml-coarse}, the DQC acts as a ``gatekeeper'' system to ensure that only reliable data are processed by the state classifier [the device state estimation (DSE) module].
The DQC module consists of a CNN classifier trained on increasingly noisy simulated data to flag data that might be of insufficient quality for reliable state estimation.
Data assessed to be of lower quality trigger one of two alternative actions: additional data analysis followed by targeted device recalibration (for data classified as of moderate quality) or tuning termination (for low quality data).
Preliminary work in this space has suggested that experimental noise and imperfections do not prevent the two-module ML system from operating as intended.
While the updated system is yet to be tested experimentally, replacing the device-dependent processing with quality assessments seems to be a step in the right direction. 
Still, there are questions that remain open in ensuring both repeatability and reliability for ML-driven automation. 

\begin{figure}[b]
\includegraphics[width=\linewidth]{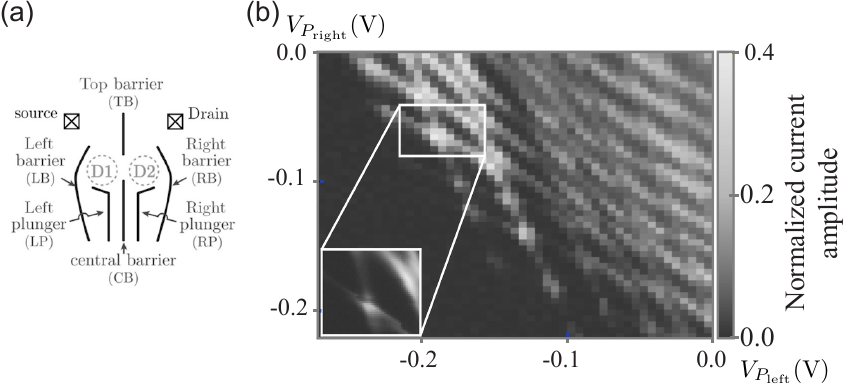}
\caption{(a) Gate arrangement of the device in GaAs used to test the tuning proposal. 
(b) Sample charge stability diagram characterized as a good double-QD regime.
The ranges to measure the 2D scan are determined based on the individual gate characterization.
The resulting scan is segmented into $0.05 \si{\volt}\times0.05\si{\volt}$ regions, which are then classified by three binary classifiers to asses the type and quality of the QD regime.
Adapted from \citet{Darulova19-ATQ}.
}
\label{fig:binary_classifier}
\end{figure}

An alternative approach to tuning device topology for double QDs in GaAs that also relies on ML was proposed by~\citet{Darulova19-ATQ}.
They set in place a sequential script-based protocol to adjust individual gates (or sets of gates) depending on the outcome of the ML module applied to a large 2D scan.
The limits of the 2D scan are determined based on the voltage ranges (i.e., the pinch-off levels and the ``widths'' of pinch-off curves) established during bootstrapping, as described in Sec.~\ref{ssec:boot}.
Once measured, the large 2D scan is segmented into small adjacent regions and each is analyzed independently by three binary classifiers trained to determine the QD regime (single or double QD) as well as its quality (good or bad); see Fig.~\ref{fig:binary_classifier}.
Depending on the outcome and the tuning goal, a scripted series of gate adjustments is then executed.
The algorithm stops when at least one segment of the charge diagram is classified as the desired QD regime.
An overall $80~\%$ success rate for tuning to a double-QD state is reported for an \emph{in situ} test using a set of five double-QD devices (from a single fabrication run) over two thermal runs.

\begin{figure}[t]
\includegraphics[width=\linewidth]{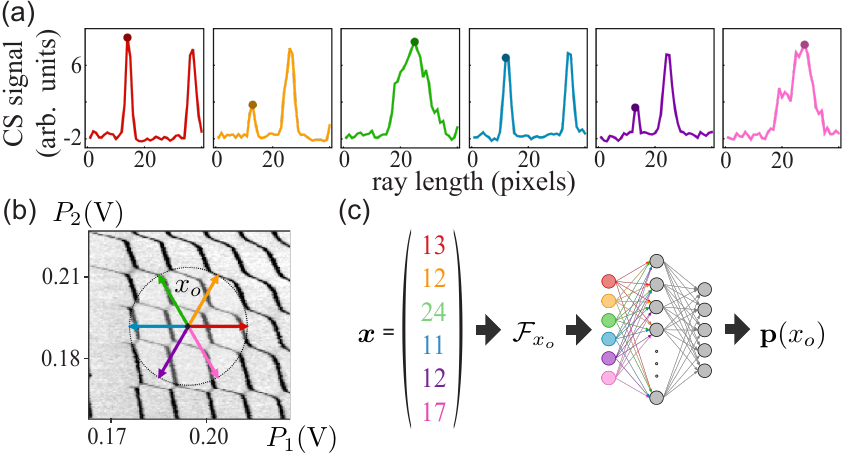}
\caption{(a) Processed charge sensor (CS) signal for six evenly distributed rays originating from a fixed point $x_o$ in a double-QD state.
In each scan, the position of a transition line nearest to point $x_o$ (that is, the ``critical feature'' along a given ray) is marked with a dot. 
(b) Distribution of rays overlaying a 2D measured scan.
(c) Visualization of the ray-based fingerprinting framework.
A vector of critical features is processed to produce a ``point fingerprint'' $\mathcal{F}_{x_o}$.
This is then analyzed using a DNN classifier, resulting in a predicted state vector ${\rm\bf{p}}(x_o)$ quantifying the current state of the device at the point $x_o$.
Adapted from \citet{Zwolak21-RBI}.
}
\label{fig:rays_cartoon}
\end{figure}

\subsubsection{Setting topology with rays}
\label{sssec:rays_coarse}
As mentioned, different configurations of the transition lines visible in a scan represent different states of the QD devices. 
Thus, while the qualitative information about the state of the device is fully encoded in the shape and orientation of those lines, the majority of the data points in any 2D scan contain background noise, which is irrelevant when setting device topology.
In fact, it has been shown that the CNN model trained to estimate the QD device's state (the DSE module in Fig.~\ref{fig:ml-coarse}) learns to ``ignore'' all of the points measured between transition lines \cite{Ziegler22-TRA}.  
One way to reduce the number of unnecessary data points collected during measurement is to employ the ray-based classification (RBC) framework developed for classifying simple high-dimensional geometrical structures \cite{Zwolak20-RBC}.

Rather than using a full 2D scan capturing a small region of the voltage space, the RBC framework relies on a collection of evenly distributed 1D sweeps (called rays) measured from a single point $x_o=(P_1,\dots,P_N)$ in multiple directions in the $N$-dimensional voltage space to assess the relative position of transition lines surrounding $x_o$.
Figure~\ref{fig:rays_cartoon}(a) shows a series of 1D charge sensor sweeps for six evenly distributed rays originating from a sample point in the double-QD state depicted in Fig.~\ref{fig:rays_cartoon}(b).
The resulting vector of distances to the nearest transition lines encodes qualitative information about the voltage space around $x_o$, effectively ``fingerprinting'' the neighborhood of $x_o$ in the voltage space.
The state of the device near point $x_o$ is determined using the point fingerprint and a simple deep neural network (DNN) trained using simulated fingerprints data; see Fig.~\ref{fig:rays_cartoon}(c).

The RBC framework has been tested using simulated data for the case of two and three QDs \cite{Zwolak20-RBC}.
It has also been implemented experimentally (both off-line and \emph{in situ}) \cite{Zwolak21-RBI} and has shown performance on par with the more-data-demanding CNN-based classification \cite{Zwolak20-AQD} while requiring up to $70~\%$ fewer measurement points.  
The off-line tuning success rate of $78.7\,\%$ is also comparable to that reported for tuning using 2D scans [$74.6~\%$; see Table I given by \citet{Zwolak20-AQD})], even though tuning employing RBC was initiated significantly further from the targeted area than tuning based on a 2D scan, with initial points sampled uniformly over a region encompassing approximately 18 compared to 9 electron transitions, respectively.
The RBC framework naturally extends to classifying convex polytopes in higher dimensions \cite{Weber21-TBR} which makes it an appealing measurement-cost-effective solution for differentiating between states of multi-QD devices.

\begin{figure}[t]
\includegraphics[width=\linewidth]{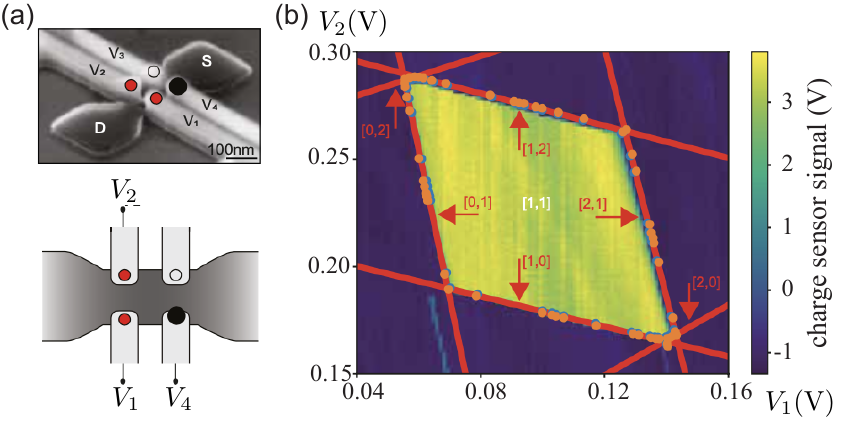}
\caption{(a) A micrograph (top image) and a schematic (bottom image) of a foundry-fabricated silicon QD device used to test the algorithm for polytope estimation.
The two red dots indicate gates used to form the qubit QDs, and the black dot represents the sensor QD.
From \citet{Ansaloni20-FFQ}.
(b) Two-dimensional map of the sensor signal as a function of the control voltages $V_1$ and $V_2$ illustrating the convex polytope of the device shown in (a), with yellow pixels indicating the $(1,1)$ state. 
The algorithm proposed estimates state transitions to other states (red lines) based on a number of point pairs $(x^+,x^-)$ (blue and orange dots) obtained via ray-based measurements.
From \citet{Krause21-ECP}.
}
\label{fig:rays-reconstruction}
\end{figure}

The RBC not only reduces the amount of data that need to be collected but also can be implemented in an online or active learning setting, where data are acquired sequentially.
In fact, an extension of the RBC framework to estimate the faces of the convex polytopes defined by state transitions was recently been proposed by \citet{Krause21-ECP}. 
They used a set of critical features measured in random directions from a fixed point $x_o$ in the voltage space combined with an algorithm for learning convex polytopes from data with a margin \cite{Gottlieb21-LCP} to approximate a candidate polytope whose faces intersect these points.
Using an active learning strategy, the polytope estimate is iteratively refined based on a consecutive set of ray-based measurements targeting the vertices of the predicted polytope until either all faces are correctly located or the computation time runs out.
The results for the double-QD system show that the algorithm can reliably find the facets of the polytope, including small facets with sizes on the order of the measurement precision; see Fig.~\ref{fig:rays-reconstruction}(b).

While the extension of the RBC framework to approximate the enclosing polytope is interesting, there are a number of questions about its practical implication that remain open.
For example, it is not clear what the advantage is of finding all of the facets over simply estimating the general class a given polytope belongs to (as originally proposed) for the purpose of tuning to a specific charge state.
Moreover, assignments of the specific transitions to a given face, such as $(2,0)$ or $(2, 1)$ in Fig.~\ref{fig:rays-reconstruction}(b), rely on the assumption that the exact current state of the devices is already known, which in practice is usually not the case when one tunes QD devices. 
Rather, the exact charge state of the QD device is typically calibrated once the QD topology is set, as discussed in Sec.~\ref{ssec:charge}. 

\subsection{Establishing controllability}
\label{ssec:control}
Once the device topology is defined, the natural next step would be to finely calibrate the device to a specific charge configuration.
As mentioned, voltages applied to the finger gates shape the overall potential affecting electrons in the 2DEG.
Ideally, changing voltages on a single gate would affect only a single parameter that the given gate is designed to control (such as the electrochemical potential of a specific QD or the tunnel barrier between two adjacent QDs).
However, owing to the capacitive crosstalk between the various gate electrodes changing the gate voltages defining one QD affects, at least to some degree, the potential of all nearby QDs.
Thus, with the growing size of QD arrays, the task of setting the charge configuration becomes increasingly challenging.

The plunger and barrier gates collectively affect the overall potential profile $\mu$, dot-specific single-particle energy detuning (i.e., the energy difference between the two QDs) of individual QDs $\delta_{ij}$, the tunnel couplings between QDs $t_{ij}$, and tunnel rates between the outermost QDs and reservoirs $\Gamma_{i}$.
One way to compensate for the capacitive crosstalk between gates is to enable orthogonal control of the QDs potential by implementing so-called virtual gates.
Specifically, linear combinations of gate voltage changes can be mapped onto on-site energy differences \cite{Oosterkamp98-MSQ}.
Virtual gates (the desired output of the establishing controllability phase) are linear combinations of multiple-QD gate voltages chosen in such a way that only a single electrochemical potential or tunnel barrier is addressed \cite{Hensgens17-FHQ, Hensgens18-PhD, Perron15-QSB}.

The effect of shifting any given physical gate on the electrochemical potentials of all other gates is typically expressed via a matrix of cross-capacitance couplings $\mathbf{C}$.
The cross-capacitance matrix can be established by finding the slopes of transition lines in 2D images measured by fixing one of the gates as a reference and then sweeping all of the remaining gates (one at a time) against the reference gate.
Alternatively, one can use the relative magnitude of a shift of steps in 1D traces of the sensing QD conductance due to changing nearby QD occupation.
At present automation of these routines (finding the numerical derivative of a feature response to a variety of gates) does not require advanced ML techniques, as it corresponds to relatively straightforward line fitting and/or peak estimation measurements. 
However, \citet{Ziegler23-AEC} recently demonstrated that these data can be extracted directly using image recognition combined with traditional fitting.

\begin{figure}[t]
\includegraphics[width=\linewidth]{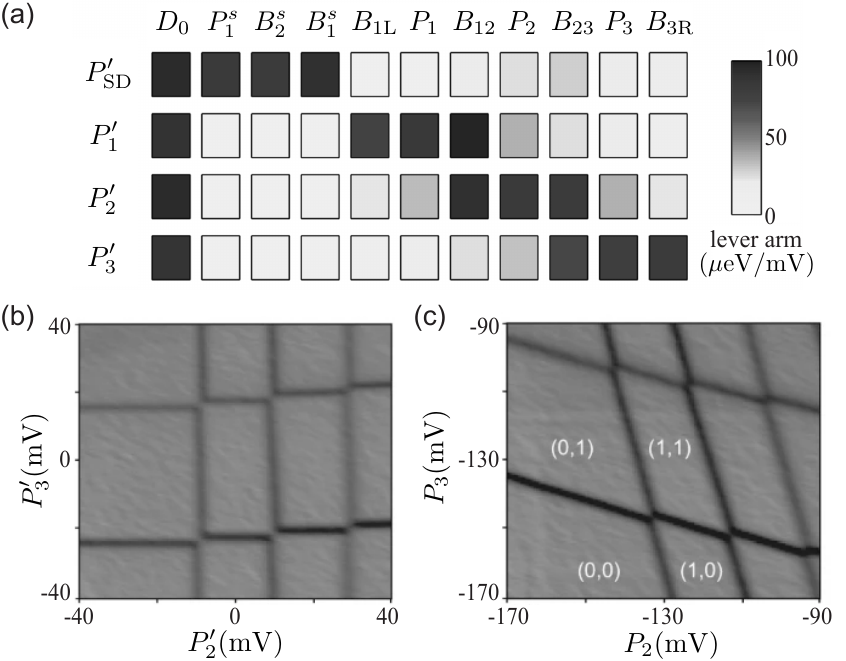}
\caption{(a) Visualization of the cross-capacitance matrix for a device shown in Fig.~\ref{fig:intro-fig}. 
The entries of each row show how the virtual plunger value (and hence the electrochemical potential) of a QD is influenced by other gate voltages. 
The rows for virtual barrier gates are omitted for simplicity. 
Gates defining the sensing QD ($P^s_1$, $B^s_1$, and $B^s_2$) are also included.
Adapted from \citet{Hensgens18-PhD}.
(b) Charge stability diagram of a double QD in the single-electron regime recorded as a function of the virtual plungers $P'_2$ and $P'_3$. 
(c) Charge stability diagram of the same double QD in the physical gates space: the gradient of the charge sensor response (in arbitrary units) as a function of the plunger gate voltages $P_2$ and $P_3$. 
From \citet{Volk19-LQR}.
}
\label{fig:virtual-gates-implementation}
\end{figure}

A visualization of the cross-capacitance matrix for a three-QD device with a one-QD sensor is shown in Fig.~\ref{fig:virtual-gates-implementation}(a).
The entries of each row in this matrix show how the physical gate voltages affect each virtual gate, with a visible falloff in gating strength with distance. 
Virtual gates $\mathbf{G}^{\rm virt}$ obtained by inverting the matrix $\mathbf{C}$ orthogonalize the voltage gate parameter space $\mathbf{G}$ and enable control of each gate independently without any concern about unwanted influence on the remaining QD energies \cite{Hensgens18-PhD}:
\begin{equation}
    {\bf G}  = {\bf C}^{-1}{\bf G}^{\rm virt}, 
\end{equation}
where $\mathbf{G}=[D_0,P_1,...,P_N,B_1,...,B_M]^T$ is a vector of physical gates defining qubit and sensor QDs and $\mathbf{G}^{\rm virt}$ is a vector of corresponding virtual gates. 
Note that some in the literature use the opposite convention, but we prefer the simple understanding that the charge space configuration is given by the capacitance matrix times the voltage, or $Q = C V$.
Figure~\ref{fig:virtual-gates-implementation}(b) shows a sample double-QD charge stability diagram measured in the virtual gate space.
For comparison, the same state measured in the physical gates space without correcting the cross capacitance is shown in Fig.~\ref{fig:virtual-gates-implementation}(c).

By enabling targeted control of specific QDs, virtual gates can be used to fill a QD array with electrons into a desired charge configuration.
However, in the original proposal the virtual gates did not account for the crosstalk on the tunnel barriers, which is necessary for fine-tuning QDs to work as qubits.
In other words, applying voltage on gate $B'_{ij}$ would affect not only $t_{ij}$ but also nearby tunnel couplings.
To correct this limitation, two proposals for a redefinition of virtual gates that compensate for the crosstalk on tunnel couplings have recently been put forward \cite{Hsiao20-EOT, Qiao20-CME}.
Both proposed corrections rely on the assumption that $t_{ij}$ can be approximated as an exponential function of a linear combination of the physical gates
\vspace{-3pt}
\begin{equation}\label{eq:exp_fit}
    t_{ij} = t_0 \exp\left(\Delta_{ij}\right) + \gamma_{ij},
\end{equation}
where $t_0$ and $\gamma_{ij}$ are the fit parameters and $\Delta_{ij}$ is a linear combination of the originally proposed virtual gates $P'_{k}$ and $B'_{kl}$, with 
\vspace{-3pt}
\begin{equation}\label{eq:Hsiao20-EOT}
    \Delta_{ij} = \sum_k\alpha_k^{ij} P'_k + 
                  \sum_{kl}\beta_{kl}^{ij} B'_{kl}, \vspace{-3pt}
\end{equation}
where $\alpha_k^{ij}$ and $\beta_{kl}^{ij}$ are derived from experimental data \cite{Hsiao20-EOT}, or 
\vspace{-3pt}
\begin{equation}\label{eq:Qiao20-CME}
    \Delta_{ij} = \sum_k\lambda_{ij} B'_{ij}, \vspace{-3pt}
\end{equation}
where $\lambda_{ij}$ are established based on a combination of simulations and experimental data analysis \cite{Qiao20-CME}. 
As mentioned, the parameters used in the fitting likely depend substantially upon the topology and number of electrons in a given QD and QD array, and thus are reliable only as part of the fine-tuning process.
\citet{Hsiao20-EOT} redefined virtual gates using ratios of the $\alpha_k^{ij}$ and $\beta_{kl}^{ij}$ coefficients from Eq.~\eqref{eq:Hsiao20-EOT}.
Similarly, \citet{Qiao20-CME} found the required virtual barrier-gate voltages necessary to achieve the targeted exchange coupling values by inverting Eq.~\eqref{eq:exp_fit}.
The resulting virtual barrier gates enable orthogonal control of the tunnel couplings over a wide range of tunnel coupling values, as shown for quadruple QDs by both \citet{Hsiao20-EOT} and \citet{Qiao20-CME}.

An alternative approach to correcting for the nonlinear and nonlocal dependence of exchange couplings on the barrier-gate voltages was also proposed and demonstrated by \citet{Qiao20-CME}.
The model developed to define the virtual gates relies on the Heitler-London (HL) expression for exchange coupling between two spins \cite{deSousa01-EMQ}.
However, solving nonlinear equations defining the HL-based model numerically can result in errors. 
Moreover, this model requires \emph{a priori} knowledge of the QD confinement potential, which makes it less desirable for practical applications.

The utility of virtual gates has been demonstrated in a number of multi-QD experiments, from showing controlled filling of an array of eight QDs using the ``n+1 method'' \cite{Volk19-LQR} and tunable coupling between single electrons in a SiMOS \cite{Eenink19-TCQ} to demonstrating shuttling of a single charge through a nine-QDs charge stability space \cite{Mills19-SSC} and transferring single-spin eigenstates and entangled states via coherent SWAP gates between neighboring pairs of spins in a four-qubit array \cite{Kandel19-CST}.
Adding the orthogonal control over tunnel couplings enabled a Heisenberg spin chain to be generated and two-, \mbox{three-,} and four-spin exchange oscillation to be demonstrated \cite{Qiao20-CME}.

\begin{figure}[b]
\includegraphics[width=\linewidth]{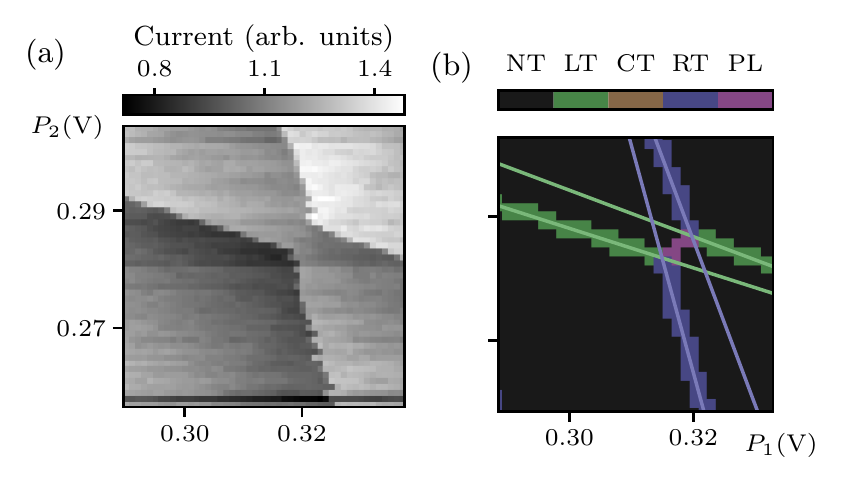}
\caption{
(a) Sample raw measurement data (b) Outcome of the ML-based pixel classifier for a sample 2D scan capturing a double-QD region.
The data were acquired using a Si/Si$_x$Ge$_{1-x}$ quadruple-QD device identical to the one shown in Fig.~\ref{fig:device-scan}.
The pixels in the original 2D scan are classified as no, left, central, or right transition (NT, LT, CT, or RT, respectively) or a polarization line (PL).
From \citet{Ziegler23-AEC}.
}
\label{fig:virtual-gates-ml}
\end{figure}

\begin{figure*}[t]
\includegraphics[width=\textwidth]{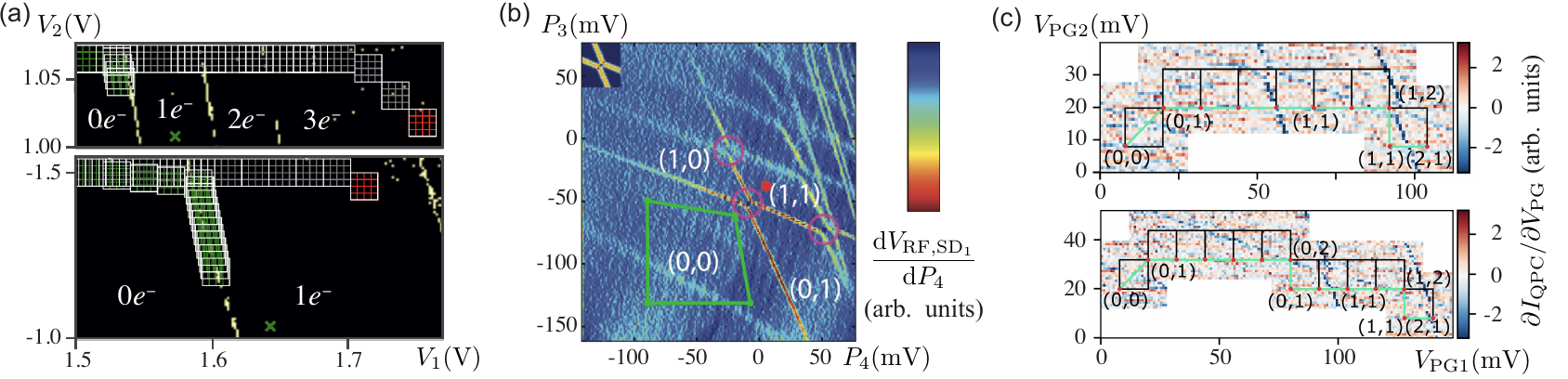}
\caption{
(a) Visualization of the patch shifting algorithm on two stability diagrams.
The red grid indicates the starting points (see also Sec.\ref{ssec:coarse}), the shifting path toward the empty reference point is shown in gray, and green denotes the shifting path toward the single-electron regime, with the green cross indicating the final position. 
After detecting the transition line from the empty reference point, a jump to the lower right is performed  to ensure that one ends up in the single-electron regime.
From \citet{Czischek21-MNA}.
(b) Sample double-QD charge stability diagram showing three good matches to the reference crossings of a charge transition lines template (inset), which are marked with purple circles.
The occupancy of each QD is denoted by $(n, m)$ and indicates the number of electrons in the left and right QDs, respectively.
The single-electron regime is found by verifying that no other charge transition lines are observed for more negative gate voltages with respect to the most bottom-left detected cross (green regions).
From \citet{Baart16-CAT}.
(c) Visualization of two successful runs of the ML-driven autotuner for tuning from a $(0,0)$ to a $(2,1)$ state.
The boxes in black denote the patches used to determine the position of a next scan.
From \citet{Durrer19-ATQ}.
}
\label{fig:tuning-charge}
\end{figure*}

More recently, \citet{Ziegler23-AEC} proposed an algorithm that leveraged a ML-based pixel classifier \cite{Lin16-FPN} and simple linear regression to determine virtual plunger gates from a small 2D measurement. 
The pixel classifier implemented in the algorithm was trained on simulated data to flag every pixel in an image as a no, left, central, or right transition (NT, LT, CT, or RT, respectively) or a polarization line (PL). 
The resulting clusters of pixels are then fit to a linear model independently for each cluster, as shown in Fig.~\ref{fig:virtual-gates-ml}. 
If multiple transitions are present within a single class, the slopes are combined by an average weighted using the standard deviations of the fits. 
The LT and RT classes yield the off-diagonal terms of the capacitive coupling matrix. 
The virtualization algorithm has been implemented in the loading module of the physics-informed tuning algorithm for charge tuning \cite{Ziegler22-TAR}. 
Its utility has been validated using both simulated and large experimentally acquired 2D scans. 
This method has the added benefit that the fits also give confidence intervals, which in turn allows for automated detection of undesired behaviors such as the formation of spurious QDs.
However, at present it does not allow virtualization of the barrier gates.

\subsection{Charge tuning}
\label{ssec:charge}
The coarse-tuning algorithms discussed in Sec.~\ref{ssec:coarse} terminate once a targeted state (such as single- or double-QD state) is reached.
However, the particular charge configuration of the system remains unknown at that point.
At the same time, various QD qubit realizations [hybrid qubits~\cite{Shi12-HSQ}, resonant exchange qubits~\cite{Medford13-REQ}, or quadrupolar exchange-only qubits~\cite{Russ18-QES}] require a specific number of electrons.

Depending on the QD qubit type, the required number of electrons is typically one to three per site.
The goal of the charge-tuning phase is to set the devices into a specific charge configuration. 
The charge sensor measurement does not provide the exact number of charges sensed, but rather the overall direction of change.
The process of determining the charge configuration of a QD device (at least for the single- and double-QD cases) typically involves counting of the transition lines while the QDs are emptied of all electrons, one at a time.
To date all proposals for automating the charge tuning have the same underlying strategy, which comprises two phases: (i) emptying the QDs of all electrons and (ii) reloading the desired number of electrons on each QD.

The two algorithms proposed for single-QD devices discussed in Sec.~\ref{ssec:coarse} implement this exact strategy.
Starting at the transition detected either via image processing \cite{Lapointe-Major19-ATQ} or using ML-driven methods \cite{Czischek21-MNA}, gate voltages are adjusted to move toward neighboring lines until no more transitions can be found.
At that point (the so-called reference point), the QD is considered emptied and a reloading phase begins.
The latter involves adjusting gate voltages in the opposite direction until the first transition line is detected, at which point the device is declared to be in a single-electron regime, as depicted in Fig.~\ref{fig:tuning-charge}(a).
The ML-driven algorithm has a high success rate of about $98~\%$ for emptying the single-QD device.
However, the reported $53~\%$ success rate for reloading a single electron back on the QD is rather low, which was attributed to interruptions in transition lines caused by the experimental measuring procedure.
To mitigate the risk of the shifting algorithm hitting a gap in the transition lines, \citet{Czischek21-MNA} proposed using arrays of $K$ by $K$ adjacent patches, as shown in Fig.~\ref{fig:tuning-charge}(a), and showed that the success rate, which seemed to strongly depend on the array size, peaks for $K=4$ at $75~\%$.

For double-QD devices, the process of counting lines is more complicated, as it is necessary to keep track of not only the total number of electrons but also which QD a given electron belongs to.
The first automation proposal for tuning double-QD devices to a single-electron regime was put forward by \citet{Baart16-CAT}.
Starting at the most bottom-left crossing detected via the pattern matching algorithm described in Sec.~\ref{ssec:coarse}, the single-electron regime is found by verifying that no other prominent charge transition lines are observed when gate voltages are made more negative, as shown in Fig.~\ref{fig:tuning-charge}(b).
At that point, the algorithm sets both plunger voltages slightly above where the most bottom-left crossing was matched to enter the single-electron regime.

An alternative, ML-driven approach to charge-state tuning, demonstrated using a GaAs triple-QD device operated in the double-DQ mode, was proposed by~\citet{Durrer19-ATQ}.
Their starting point for the charge-tuning algorithm was a well-defined double-QD state with an unknown charge state $(n_i,m_i)$, where $n_i$ and $m_i$ denote the unknown initial numbers of charges on each QD, which is precisely the end point of the coarse-tuning phase discussed in Sec.~\ref{ssec:coarse}.
Then, a two-stage algorithm for tuning the double-QD device to a preselected charge configuration $(n,m)$ following the unloading--reloading strategy is initiated. 
Both stages involve a series of measurements in the space of the plunger gates, with each followed by a CNN analysis of the measured scan and appropriate voltage adjustments rooted in heuristics.
The measurement analysis is carried out by two specialized CNNs, each pretrained using about $10^5$ images sampled from large, experimentally measured stability diagrams that are manually labeled with the position and orientation of the charge transition lines, with additional charge stability diagrams obtained by applying multiple augmentations to the experimental scans.

For the first (unloading) phase, a small ($20 \times 20$ pixel), low-resolution scan (ranging from 6 to 9\,\si{\milli\volt} per pixel, depending upon the scan, but fixed in pixel resolution) is analyzed by a binary CNN classifier for the presence of transition lines.
The size of the scans is chosen to ensure that even within the voltage range with the largest line spacing between consecutive transitions two transition lines can be captured.
The algorithm follows a fixed path, decreasing both plunger gate voltages whenever at least one line is detected.
Once a ``no lines'' class is identified, the depleting step terminates and a $(0,0)$ reference point charge state is assumed. 
At this point the second (reloading) phase, aimed at loading electrons back to achieve the desired charge state, is automatically initiated.

The second algorithm also follows a predefined, although admittedly more nuanced, path.
In particular, as long as both QDs need more electrons, both plunger gate voltages are increased by a fixed amount.
Once at least one of the QDs reaches the desired number of electrons, the direction of consecutive steps depends on which of the QDs needs to be loaded (resulting in an increase of the voltage only on the relevant plunger gate) and whether any undesirable transition lines are encountered for a QD that was previously tuned (resulting in a decrease of the voltage on the appropriate plunger gate).
Two paths for the reloading phase are shown in Fig.~\ref{fig:tuning-charge}(c), where the device is tuned to a $(2,1)$ charge state. 
A separate CNN trained to recognize the presence and orientation of the detected transition lines from small ($28 \times 28$ pixel) high-resolution scans (1\,\si{\milli\volt} per pixel) is used in this phase; this higher resolution is used to get a slope, as we no longer need to see multiple charge transitions.
This time the size of scans is chosen to assure that at most one transition line can be captured.

This charge-tuning algorithm has been validated experimentally using two double-QD systems within the same triple-QD device in GaAs.
With the targeted state chosen randomly from a set of four possible configurations, an overall tuning success rate of $56.9\,\%$ has been reported for on-line tuning, with individual success rates for the unloading and reloading phases being $90\,\%$ and $63\,\%$, respectively.
The majority of errors for the second phase stemmed from the low signal-to-noise ratio resulting in the CNN missing transition lines present in the data and ending up with the addition of too many electrons. 
The unloading-reloading strategy has also been implemented by \citet{Ziegler22-TAR}.
However, rather than using 2D scans, the charge-tuning algorithm [called physics-informed tuning (PIT)] takes advantage of the ray-based measurements \cite{Zwolak20-RBC} combined with virtual gates \cite{Hensgens17-FHQ, Hensgens18-PhD} to navigate the voltage space. 
The success rate for charge setting with PIT when testing with simulated data is $95.5(5.4)~\%$ and $89.7(17.4)~\%$ (median $97.5~\%$) for off-line experimental tests.
This high performance is demonstrated on data from samples fabricated in both an academic clean room as well as on an industrial 300 \si{\milli\meter} process line, further underlining the device agnosticism of PIT.
This work presents an important step toward autonomous QD calibration by enabling the automated transition to fine-tuning discussed in Sec.~\ref{ssec:fine}.

\subsection{Fine-tuning}
\label{ssec:fine}
Once the QD system is in the desired configuration (in terms of both topology and charge), there remains a significant additional step before the system can be used as a collection of quantum bits. 
These efforts entail fine-tuning of the parameters to enable the system to be treated in an abstract manner. 
There are two main aspects to fine-tuning. 
One is ensuring that the system is behaving close to the qubit regime. 
The other is ensuring that qubit performance and controllability can be achieved. 
As discussed, we do not cover anything in the latter category, as it remains nascent in the quantum dot community, but note that other works in superconducting qubits are largely applicable to many of these tasks \cite{Kelly18-QDG,Klimov20-SOQ}.
For this work fine-tuning begins the process of mapping experimental analog parameters to those of qubits, and thus represents a transition in which information is revealed or gained by experiments.

One critical challenge for ensuring qubit behavior is making certain that charge transitions occur only when desired. 
For example, during the manipulation of electron spins as quantum bits, tunneling of an electron into a nearby Fermi sea would destroy coherence and would be considered a ``leakage'' error, in which case the qubit itself has been lost \cite{Elzerman04-SRE}.
More generally, controlled electron tunneling to a nearby Fermi sea forms a critical piece of the quantum control approach for resetting, and in some cases measuring, the qubits built from the electrons. 
In single-spin readout, a common approach put forward by \citet{Elzerman04-SRE} relies upon a good understanding of the spin-dependent tunneling of electrons into the Fermi sea. 
Alternatively, in singlet-triplet qubits, ejection of electrons followed by a refill is used to prepare initial electron singlet states. 
Thus, reservoir coupling is important to calibrate and tune, but its importance also depends upon the approach to qubit realization.
This becomes nontrivial when the additional requirement of being able to also reset or refill the QD from such reservoirs is imposed, as occurs in a variety of qubit preparation and readout schemes.

Several groups have investigated the means of automated estimation of tunneling to leads, including via the estimation of gate-dependent cotunneling rates and by observation of charge loss and reset dynamics \cite{Botzem18-TSD}. 
These approaches take advantage of charge-based measurements, such as that provided by a nearby QD or quantum point contact in use as a charge sensor.
However, rather than working (effectively) in a low frequency or dc domain, they use time dynamics. 
Analogous to estimating the time constant of an $RC$ filter or the ringdown of a mechanical resonator, observation and fitting of such curves add additional depth to the challenge. 
This depth could be addressed in an automated manner via ML techniques for processing time-domain signals but so far remains untreated.

Another challenge arises when the topology and charge number is correct but the tunneling rates to leads cannot get to the regime of interest, which may arise from pinch-off of a barrier. 
In that setting, returning to the coarse-tuning mode with a new starting point may enable one to find a different regime of parameter space that does not have such problems. 
However, no generic restart method has yet been proposed or implemented in experiments.

\begin{figure}[b]
\includegraphics[width=\linewidth]{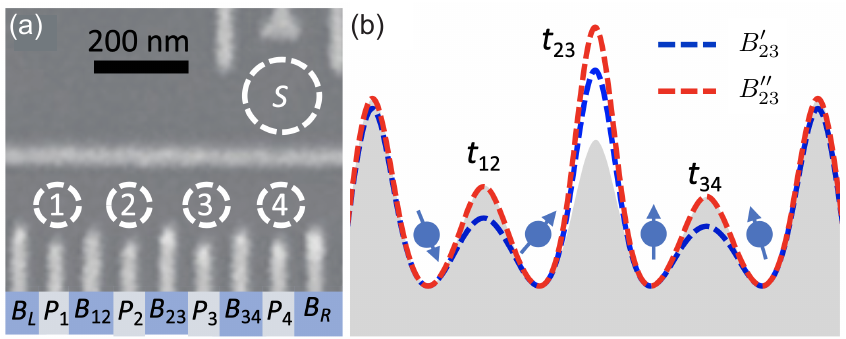}
\caption{
(a) Electron micrograph of a quadruple-QD array in GaAs.
(b) Schematics illustrating the influence of changes in virtual gates on the potential landscape of a quadruple QD.
The gray area denotes the original landscape, the blue dashed line indicates the landscape when virtual gates are defined as proposed by \citet{Hensgens18-PhD}, and the red dashed line indicates the landscape when virtual gates also compensate for the crosstalk on tunnel couplings.
As can be seen, although $B'_{23}$ keeps the QD potentials fixed  while controlling the interdot tunnel coupling $t_{23}$, it also influences $t_{12}$ and $t_{34}$. 
In contrast, $B''_{23}$ does not affect the other two interdot tunnel couplings.
From \citet{Hsiao20-EOT}.
}
\label{fig:virtual-gates-cartoon}
\end{figure}

Finally, additional errors can occur in certain materials, such as silicon and SiGe QDs. 
In those devices, in addition to the spin and orbital physics of the QDs, there is also a valley degeneracy that plays the role of a pseudospin. 
This degeneracy is typically lifted due to various interface effects or disorder in a QD device. 
In practice, many groups work to ensure that QDs have valley splitting well in excess of the temperature and of the Larmor precession period of the spin degrees of freedom, such that it effectively can be neglected. 
However, for automated tuning, valley splitting must be estimated, perhaps using tunneling spectroscopy \cite{Shi12-HSQ, Zajac16-SGA, Dodson21-VSS, Elzerman04-ESS}. 
While this in itself is easy to automate, the techniques necessary to move into a regime of the desired valley splitting while maintaining topology, electron number, and controllability has yet to be automated and may face unexpected challenges, not the least of which is that the device may not have access to the desired valley splitting at all.

The previously mentioned techniques are designed to ensure that the QD system has an appropriate means of resetting spins and does not ``leak out'' of the desired physical subspace, typically that of spin-1/2 electrons or combinations of spin-1/2 electrons. 
However, operation on the spins or qubits (for composite spin systems) still requires substantial calibration. 
One key parameter is the tunneling energy between two adjacent QDs, which determines the exchange interaction and forms the core of two-qubit gates and, for multielectron qubits, single-qubit gates as well.
This parameter can be estimated via more traditional gate voltage sweeps and by more complex time-domain investigations. 
It does lead to a challenge for automated tuning, as the rate is often exponentially sensitive to applied voltages.
Moreover, owing to the finite screening of nearby gate voltages, the tunneling energy has substantial interdependencies with other gates.

In one of the first works addressing automated control and adjustment of tunneling rates, a script-based method was developed that leveraged the virtual gate concept \cite{vanDiepen18-ATC}. 
In that effort, two different means of extracting the tunnel coupling were considered, both with low-frequency measurements (width of the charge transition) and with high-frequency measurements (photon-assisted tunneling). 
The benefit of the latter technique is the ability to measure beyond the thermal limit of linewidth, a necessity for estimating small parameters even in a dilution refrigerator setting.

The use of virtual gates described in Sec.~\ref{ssec:coarse} can help with isolating control of the tunneling rate \cite{Hsiao20-EOT,Qiao20-CME}.
The schematic in Fig.~\ref{fig:virtual-gates-cartoon}(b) depicts the effect of changes in the virtual gates (linear combinations of voltages applied to physical gates) on the potential landscape of the quadruple QD shown in Fig.~\ref{fig:virtual-gates-cartoon}(a).
Changing voltage on virtual gate $B''_{23}$ affects the targeted interdot tunnel coupling, as indicated by the red dashed line.

A different approach for handling parameter estimation is to use template matching. 
In this approach, particular shapes and characteristics of data traces are assumed, and optimization occurs via fitting real data to a set of templates and extracting effective model parameters through this mechanism \cite{Mills19-CAT}. 
A related set of techniques has also been leveraged to estimate, and tune, parameters in tunneling that lead to so-called bias triangles, using autoencoders to reduce the effective information coming from the experimental system to a significantly smaller feature space \cite{vanEsbroeck20-FTQ}. 
A key open question in fine-tuning remains in determining how much actual experimental information should be extracted at each step.
Experimental images are expensive (in time) to take, and thus reduced data acquisition is desired where possible. 
The success of autoencoders suggests that there is a latent data space that would require substantially fewer experimental data points to populate.

As the tunneling rate becomes set, additional spin parameters must then be extracted, and in some cases tuned. 
These include both the Zeeman splitting (Larmor precession frequency) and, where appropriate, estimates of the magnetic field gradient seen in each QD.
In some sense, the goal of the physics-domain fine-tuning is often to populate parameters in an effective model of the QD system. 
We note that the capacitive model used for virtual gates already provides a good starting point for this model.

Other, more phenomenological models such as the charge transition width (a proxy for tunnel coupling) and lead tunnel rates can be fit using data. 
If these estimates and their derivatives can be tracked, it becomes relatively straightforward to search parameter space efficiently.
However, tracking derivatives using finite differences becomes expensive in terms of data acquisition. 
\citet{Teske19-MFT} removed this complexity by use of a Kalman filter to enable effective characterization of the gradients, thereby enabling tuning in an efficient (in time and memory) control system.

Thus far measurement methods have generally yielded relatively direct information regarding issues such as changes in charge as a gate voltage is moved. 
However, future work must address the transition from a physics model description, such as the Hubbard model, and quantum gate performance. 
This effort need not rely directly upon experimental systems, as modeling of the system from a physics perspective is relatively straightforward given the observed connection between actual gate voltages and model parameters. 
For example, ML subsystems trained on simulation data were able to work with little degradation of performance, including typical sources of experimental noise \cite{Ziegler22-TRA}. 
However, hysteresis (for instance due to charge locking) remains a potentially challenging example of real experimental imperfections that require one to move to an online learning format.
A general solution of finding good quantum gates given this connection remains to be investigated.

\section{Outlook: Solving the metaproblems and paths forward}
\label{sec:outlook}
Thus far we have considered the scenario in which the number of QDs and the number of gates to be manipulated, as well as the number of readout systems, are kept relatively small. 
The path to scaling up spin qubit systems through these automated approaches remains open but is also opaque. 
Thus, we now consider several metaproblems that will require substantial effort before scale-up can be achieved. 
For simplicity, we focus on the four most pressing today: benchmarks and standards for assessing tuning success and performance, improving qubit yield and homogeneity, adding additional QDs to a tuned system in an efficient manner (scale-up), and moving key aspects of the ML systems closer to the QDs themselves (reducing fan-out).

\subsection{Benchmarks and standards}
\label{ssec:ben-and-tan}

A simple but crucial component of success for the field will be to solidify key metrics of performance and key datasets that can be used to assess those metrics.
This standardization work will take time and community engagement based upon experience from other ML disciplines. 
However, once standardization is in place, more algorithmic exploration and improvement can be achieved. 
In this Colloquium a few simple metrics have been suggested. 
One is state identification accuracy (the probability of a classifier identifying the right topology); another is tuning success (the probability of an optimizer getting to the right region of parameter space). 
There are many more such metrics, and associated datasets, that will be necessary to leverage ML algorithms most effectively.

To date a reasonable fraction of training and assessment has been done using simulated device data [see~\citet{qf-data}], which has the benefit of being prelabeled and being inexpensive to generate. 
However, these simulation tools are not equipped to cover all aspects of tune-up and do not capture many realistic experimental imperfections. 
A more ideal approach would be to combine synthetic and experimental datasets for each key task outlined in the tune-up process with associated performance metrics and human performance assessments to enable the field to make substantive progress.

\subsection{Improving qubit homogeneity}
\label{ssec:homogen}
To achieve scaling of QD systems we anticipate that improvements will be necessary in both automated tuning and the yield and homogeneity of QD systems themselves. 
As an initial target, we consider the effective combined yield (tuning and fabrication) to realize a surface code patch of $5\times5$, corresponding to 49 qubits. 
This suggests that $98~\%$ yield is a likely starting point for scaling, and is consistent with today’s superconducting array processors, which at 50--100 qubits typically have one to two that are not fully operational. 
Achieving this starting point remains a substantial challenge in the field, although recent work at Delft University of Technology reached 16 QDs with appreciable yield \cite{Borsoi22-QCA}.

A path toward automated lithographic quality assessments of QD devices has also recently been put forward.
The proposed control systems rely on automated probing at room temperature \cite{Zwerver22-QMA} or use CNNs applied to scanning electron microscope (SEM) micrographs collected in-line to assess the device usability based on detection of certain fabrication defects (such as particle contamination and proper exposure) \cite{Mei20-OQF}.   
However, while these are important steps toward streamlining QD device fabrication quality control, there are a number of imperfections that still need to be accounted for.
These include variations in a single-QD device over time, between cooldowns, and between changes in the control systems and other aspects of the integrated experimental system. 
This naturally fits into an online learning model, in which the running experiment is continually changing.
At the same time, the practical capabilities that subsystems that are pretrained provide may need adaptation or replacement once these concerns are integrated.

As standardization and qubit fabrication improve, we anticipate that reliably tuning up a large number of devices to the desired topology and electron number may become fully automated. 
However, this does not yield qubits by itself. 
Specifically, additional work is necessary to bring the gap between qubit tuning techniques used in quantum computing (randomized benchmarking, gate set tomography, etc.) and the physical implementation of key quantum operations that rely upon the results of fine-tuning. 
These include initialization of the qubit state, readout of the qubit state, understanding and characterizing leakage from logical states to other states of the system, and understanding and calibrating logical gate operations. 
This is an area ripe for future exploration.

\subsection{Scaling: Inductive and {\it in situ}}
\label{ssec:scaling}
On the scale-up problem, we have already seen that moving from physical gates to virtual gates is a significant benefit for controllability and fine-tuning. 
Consider now a 1D array of gates on a nominal nanowire that could be used to form $N$ QDs in series, with a nearby set of charge sensors. 
What approach can we use to ``add'' an additional QD to the array if we have already tuned up the first $k$ QDs and developed their virtual gate representation? 
Put simply, we can attempt to add the next QD holding the $k$-QD system solved and ``resolving'' the $k$th QD in conjunction with the next QD in the line, thus mapping the system to a double-QD tuning problem, but now starting with virtual gates for the $k$th QD such that crosstalk to the $k-1$ QDs is close to zero. 
A preliminary example of this inductive approach, albeit without automation, was given by \citet{Volk19-LQR}.

A different approach would instead be to tune each pair of QDs (1 and 2, 2 and 3, etc.) while leveraging virtual gates from all the prior solutions. 
One can then hope that the full solution is close to the tuned system. 
Many QD experiments with large arrays have used variations of this {\it in situ} approach to refine the range of acceptable parameters for searching for QDs. 
One advantage of this approach is the ability to keep both the left and right QDs in contact with the Fermi sea, making charge trapping and other hysteretic effects negligible.

In both cases, the assumed 1D nature of the physical layout provides substantial simplifications. 
More complicated is the loading of a 2D array of QDs, although both approaches may also find success in that setting. 
We do not anticipate that a fully computerized system is yet available that can explore this space efficiently. 
Specifically, handling both the challenges of charge latching (metastable states) previously described for 1D systems and the difficulties of assessing sensing knowledge from a number of sensors that grows only as the boundary (and gets further from the actual QDs) remains an open problem.
Furthermore, the actual state space is growing exponentially, but the space of interest is much narrower. 
Teaching ML systems to find and stay within this narrow regime is an open problem. 
Thus, the path forward today remains adapting practice in the lab to ML systems for solving key subproblems first, then looking at generalizations.

In most of this Colloquium, we have focused on relatively stable devices that are generally ``well behaved,'' that is, that work within the expectations of the trained systems described here. 
However, real-world systems and devices have a variety of imperfections that will need to be considered as systems move toward scaling up and integration. 
Two specific examples of concern are handling spurious QDs (regions of parameter space in which the topology changes due to a defect or impurity that acts as an additional charge trap) and handling latching (metastable states) and related hysteresis effects.
Latching occurs when the lowest energy charge configuration cannot be reached as the electrons necessary to reach that configuration cannot get to the right region of the device. 

Thus, in addition to finding a path to scaling up, we suggest that additional efforts also consider the use of techniques common in the online learning setting to handle unexpected behaviors. 
For example, reinforcement learning techniques can be used for optimization in the hysteretic setting associated with charge latching, as the optimizer has a sense of the history of the measurement (much like navigating a maze), rather than just the current state of the system. 
Similarly, online learning approaches can provide sandboxing or related techniques to identify and then stay away from regions with undesired behavior (be it too noisy, too weak for measurement, or due to spurious QDs).
Preliminary work on this is ongoing \cite{Ziegler22-TRA,Ziegler23-AEC}.

\subsection{Toward ``on-chip'' implementations}
\label{sec:on-chip}
Finally, with the noisy intermediate-scale quantum technology
era on the horizon \cite{Preskill18-QCN}, it is timely to consider the practical aspect of implementing autonomous control.
Present-day devices in which quantum bits are stored and processed require classical electronics to measure and control the qubits, as well as conventional computer software to control and program these electronics. 
Since each qubit must be controlled and measured separately, it is necessary for the classical control system to scale along with the number of qubits, which at present represents a substantial engineering challenge  \cite{Vandersypen17-ISQ,Geck19-CEQ}, particularly as systems scale into the thousands or millions of qubits regime.

While a complete, ``in fridge'' solution may be of interest, we note that a variety of key subtasks may be the only ones necessary to achieve high-throughput tune-up and calibration of systems in practice at this scale. 
Techniques such as ray-based learning can currently reduce dataflow requirements and thus improve performance \cite{Zwolak20-RBC, Chatterjee21-AEC}. 
Other simple recognition tasks in measurement systems may be a natural next point for improvement, in which real-time filtering and processing can be handled close to the QDs, potentially enabling more rapid calibration, particularly in the fine-tuning stage.

Placing the qubit control and analysis electronics in the immediate vicinity of the qubits in the cryostat either on the same chip or through chip-to-chip interconnection technology poses serious constraints in terms of sizing and energy requirements. 
Thus, computational requirements and power consumption should also be considered when one designs the autotuning system, enabling miniaturization of the control elements on low-power hardware, which is a significant step toward on-chip autotuning.
This issue has recently been discussed in the context of identification of charge-state transition lines in 2D stability diagrams, which is an important component in many autotuning proposals \cite{Czischek21-MNA,Zwolak21-RBI}.

\citet{Czischek21-MNA} recently showed that an extremely small feed-forward neural network (FFNN) with just a single hidden layer can be trained to detect charge-state transitions in single-QD stability diagrams.
Using a dataset of 800 synthetic stability diagrams \cite{Genest20-MSc}, they trained a binary classifier with only ten hidden neurons to differentiate between the small patches with and without transition lines and reported a testing accuracy of $96~\%$ on patches sampled from experimental data.
The FFNN classifier lies at the core of a ``shifting algorithm'' \citet{Czischek21-MNA}  proposed for tuning single-QD devices to the single-electron regime; see Sec.~\ref{ssec:coarse} and \ref{ssec:charge}.

An alternative approach to miniaturizing ML models was proposed by \citet{Zwolak21-RBI}.
Here, by replacing the traditional 2D scans and a CNN-based classifier with a series of 1D rays and encoding the state of the device via fingerprints (see Sec.~\ref{ssec:coarse}), the total number of parameters defining the ML model used for state assessment was reduced by 2 orders of magnitude compared to a CNN model ($1.2\times10^4$ versus $2.2\times10^6$ parameters); see \citet{Zwolak20-AQD}).
While this number is still fairly high, it has been noted that the size and complexity of the DNN can be further reduced by at least an order of magnitude \cite{Zwolak20-RBC}.
The RBC does not require any significant data processing, which further improves the computational efficiency of this approach and makes it an appealing candidate for the on-chip implementation on miniaturized hardware with low power consumption \cite{Sebastian20-MIC}.

\section{Conclusions}
\label{sec:concl}
With these provisos in mind, we note that there is an expansive set of tools that are being explored and leveraged to make QD systems substantially easier to operate and more reliable in their execution.
The efforts described here are primarily targeted at the initial elements of tune-up, rather than those of fine-tuning for qubit operation. 
Substantial efforts in superconducting and ion-based quantum computing systems have already led to the development of robust methods for such fine-tuning \cite{Arute20-ODF,Maksymov21-OCG,Gerster22-EBC,Majumder20-RTC}. 
In addition, a key data-based approach, the use of directed acyclic graphs, can be a natural extension of the existing script-based methods used for the tuning described in this Colloquium \cite{Kelly18-QDG}. 
However, this has not yet been integrated into workflows in today’s QD experiments, in part due to the software development and engineering requirements to fully take advantage of these advances.

Some portion of the failures observed in autotuning are attributable to the initial starting point for tuning, and thus a relevant but underexplored parameter is the success probability given a ``repeat-until-success'' tuning approach, which has not yet been studied in experimental systems. 
However, even with these improvements, we anticipate that substantially better accuracy will be required to tune large QD systems in a fully automated fashion.

Whether materials science and fabrication can reduce the need for this automation remains a question of both science and engineering. 
Practically speaking, the critical question will be one of the cost, in time and effort, to leverage ML for key tasks when compared to less complex methods.
We suspect that automated systems of this nature will be of critical importance for enabling ever-more-complex experiments for the foreseeable future.

\begin{acknowledgments}
The views and conclusions contained in this Colloquium are those of the authors and should not be interpreted as representing the official policies, either expressed or implied, of the U.S. Government. 
The U.S. Government is authorized to reproduce and distribute reprints for Governmental purposes notwithstanding any copyright noted herein. 
\end{acknowledgments}

%
\end{document}